\documentclass[trackchanges,twocolumn]{aastex701}

\newcommand{\oi}{\hbox{[O\,{\sc i}]}}
\newcommand{\ha}{\hbox{H$\alpha$}}
\newcommand{\hb}{\hbox{H$\beta$}}

\newcommand{\oiii}{\hbox{[O\,{\sc iii}]}}
\newcommand{\nii}{\hbox{[N\,{\sc ii}]}}
\newcommand{\sii}{\hbox{[S\,{\sc ii}]}}

\newcommand{\hi}{\hbox{H\,{\sc i}}}

\begin{document}

\title{Jet-ISM Interaction and Multi-channel AGN Feedback in the Post-merger Galaxy 4C+29.30}

\author[orcid=0000-0002-8411-3314]{Xiao Cao}
\affiliation{Department of Astronomy, Xiamen University, Xiamen, 361005, People's Republic of China}
\email{caoxiao24@xmu.edu.cn}  

\author[orcid=0000-0003-4874-0369]{Junfeng Wang} 
\affiliation{Department of Astronomy, Xiamen University, Xiamen, 361005, People's Republic of China}
\email[show]{jfwang@xmu.edu.cn}

\author[orcid=0000-0003-3226-031X]{Yan-Mei Chen}
\affiliation{School of Astronomy and Space Science, Nanjing University, Nanjing 210023, People's Republic of China}
\affiliation{Key Laboratory of Modern Astronomy and Astrophysics (Nanjing University), Ministry of Education, Nanjing 210023, People's Republic of China}
\email{chenym@nju.edu.cn} 

\author[orcid=0009-0005-9342-9125]{Min Bao}
\affiliation{School of Physics and Technology, Nanjing Normal University, Nanjing 210023, Peoples Republic of China}
\affiliation{School of Astronomy and Space Science, Nanjing University, Nanjing 210023, People's Republic of China}
\affiliation{Key Laboratory of Modern Astronomy and Astrophysics (Nanjing University), Ministry of Education, Nanjing 210023, People's Republic of China}
\email{mbao@nnu.edu.cn} 

\author[orcid=0000-0003-0970-535X]{Xiaoyu Xu}
\affiliation{School of Astronomy and Space Science, Nanjing University, Nanjing 210023, People's Republic of China}
\affiliation{Key Laboratory of Modern Astronomy and Astrophysics (Nanjing University), Ministry of Education, Nanjing 210023, People's Republic of China}
\email{xuxy95@nju.edu.cn} 

\author[orcid=0009-0007-7542-1140]{Chunyi Zhang}
\affiliation{Department of Astronomy, Xiamen University, Xiamen, 361005, People's Republic of China}
\email{1412133683@qq.com} 

\author[orcid=0000-0002-2853-3808]{Taotao Fang}
\affiliation{Department of Astronomy, Xiamen University, Xiamen, 361005, People's Republic of China}
\email{fangt@xmu.edu.cn}

\author[orcid=0000-0001-7349-4695]{Jianfeng Wu}
\affiliation{Department of Astronomy, Xiamen University, Xiamen, 361005, People's Republic of China}
\email{wujianfeng@xmu.edu.cn}

\begin{abstract}
4C+29.30 is a post-merger galaxy hosting a rejuvenated active galactic nucleus (AGN) with a complex multi-scale radio morphology, making it an ideal laboratory to study the interplay between different AGN feedback modes. We present a multi-wavelength analysis combining optical integral field spectroscopy (SDSS/MaNGA and CFHT/SITELLE) with radio continuum imaging (VLASS) to map the ionized gas kinematics and ionization structure across the galaxy. We uncover a galaxy-scale, biconical ionized gas outflow whose axis is misaligned by $\sim$26$^\circ$ from the radio jet. This outflow, characterized by broad line widths and Seyfert-like ionization, is mostly consistent with a radiatively driven wind from the central supermassive black hole, which is accreting at a relatively high Eddington ratio ($L_{\mathrm{bol}}/L_{\mathrm{Edd}} \gtrsim 0.1$). In contrast, the northern radio lobe clearly drives localized gas acceleration and increased velocity dispersion, indicative of jet-driven shocks interacting with the interstellar medium, consistent with previous X-ray findings. The coexistence of a radiatively driven galactic-scale outflow and a distinct, misaligned radio jet demonstrates that multiple AGN feedback channels can operate simultaneously within the same system, providing new evidence for the concurrent action of radiative and mechanical feedback.
\end{abstract}


\section{Introduction} \label{sec:intro}

The energy released by active galactic nuclei (AGNs), powered by accretion onto central supermassive black holes (SMBHs), can profoundly influence both the interstellar medium (ISM) and the surrounding intergalactic medium (IGM) \citep{Fabian2012,Heckman2014}. Through the injection of energy and momentum, AGNs can drive multi-phase outflows that extend from galactic nuclei to circumgalactic scales. Consequently, the cold gas reservoir fueling star formation may be expelled or heated, thereby suppressing star-forming activities (negative feedback, \citealt{Birzan2004,Croton2006,McNamara2012,Somerville2015}). Alternatively, AGN-driven compression of gas can enhance star formation (positive feedback, \citealt{Gaibler2012,Mukherjee2021}). In this way, AGN activity plays a significant role in host galaxy evolution by regulating gas kinematics and mass assembly history.

Observationally, local AGNs are commonly divided into two principal feedback modes \citep{Heckman2014}. The first consists of systems in which the dominant energy output is electromagnetic radiation produced by efficient accretion onto the SMBH. These are referred to as radiative- or quasar-mode AGNs and are typically characterized by relatively high Eddington ratios ($\lambda_{\rm Edd} \gtrsim 0.01$). Radiative-mode feedback is commonly associated with outflows. Numerous spectroscopic studies have revealed blueshifted absorption features in the optical, ultraviolet, and soft X-ray bands, indicating outflow velocities ranging from several hundred km s$^{-1}$ in low-luminosity AGNs to tens of thousands of km s$^{-1}$ in luminous quasars \citep{Chartas2002,Pounds2003,Reeves2003,Crenshaw2003,Blustin2005,Tombesi2010}. Over the past two decades, integral-field spectroscopy (IFS) observations have further identified kiloparsec-scale biconical outflows in obscured (Type 2) AGNs, often oriented perpendicular to the host galaxy disk (e.g. \citealt{Bao2019,Juneau2022,Zhou2024,Marconcini2025}). The most powerful radiative-mode AGNs are now widely considered capable of driving multi-phase outflows that substantially affect their host galaxies (e.g. \citealt{Maiolino2012,DallAgnoldeOliveira2023}), although the net impact on star formation remains under active debate \citep{Cresci2015a,Cresci2015b,Maiolino2017}.

On larger scales, direct evidence for AGN feedback is most clearly observed in massive elliptical galaxies. In these systems, radio jets inflate cavities or bubbles in the hot halo gas, depositing mechanical energy as the relativistic plasma-filled lobes expand. This process can offset or delay the cooling of the surrounding hot gas \citep{Birzan2004,McNamara2012}. Such feedback is typically referred to as radio-, jet-, or mechanical-mode AGN activity and is associated with inefficient accretion at low Eddington ratios ($\lambda_{\rm Edd} \lesssim 0.01$). In addition to affecting the CGM, jets can strongly interact with the ISM of the host galaxy before propagating to larger scales. Numerical simulations of jet-ISM interactions demonstrate that radio jets can significantly redistribute gas content and potentially regulate star formation (e.g., \citealt{Mukherjee2021}). Understanding jet-ISM interactions is therefore crucial for assessing how mechanical feedback operates within galaxies. 

While radiative and mechanical feedback modes are often discussed separately, a subset of high accretion-rate AGNs can launch powerful radio jets, such as radio-loud quasars (e.g. 3C 273, \citealt{Husemann2019a}) and high-excitation radio galaxies (HERGs; e.g. Cygnus A, \citealt{Tadhunter2003}), indicating that radiatively efficient accretion and powerful jets can coexist within a single AGN phase. However, the relative roles of radiative and mechanical feedback within individual radio-bright AGNs remain poorly constrained.
In radio-bright AGNs, mechanical feedback is often assumed to dominate, and ionized gas outflows are therefore commonly attributed to jet-ISM interactions \citep{Mullaney2013,Jarvis2019}. It remains uncertain whether radiatively driven galactic-scale winds contribute significantly. Distinguishing these mechanisms is particularly challenging in systems where prominent radio structures coexist with signatures of active nuclear accretion. Spatially resolved observations with a sufficiently large field of view (FoV) are therefore essential to disentangle their relative impacts.

A post-merger galaxy 4C+29.30 (also referred to as B2 0836+29A, \citealt{OcanaFlaquer2010}), with a moderate radio luminosity of $\sim 10^{42}\ \rm erg\ s^{-1}$ \citep{Siemiginowska2012}, exhibits galactic-scale radio structures including core, jets and double lobes, in which the radio major axis is almost aligned with the galaxy disk. Naturally, 4C+29.30 provides an excellent laboratory for investigating AGN feedback and, in particular, the nature of jet-ISM interactions.  
Early radio and  optical studies revealed a close spatial correspondence between the radio emission and optical line-emitting gas, suggesting strong coupling between the radio source and its environment \citep{vanBreugel1986}. Subsequent radio observations uncovered radio structures with different evolutionary timescales, indicating episodic jet activities from pc- to kpc-scales \citep{Jamrozy2007,Liuzzo2009}. Redshifted \hi\ absorption detected toward the nucleus further reinforced the possibility that infalling cold gas may be fueling the recent black hole activity, likely originating from a gas-rich merging process \citep{Chandola2010}.

Multi-wavelength observations for 4C+29.30 have revealed correlations between both hot and warm ionized gas and the radio structures. X-ray observations with Chandra indicate that thermal gas surrounding the northern radio lobe may be shock-heated by the expanding radio source \citep{Siemiginowska2012}. Optical IFS observations with Gemini further identified high-velocity and high-dispersion nuclear gas consistent with a bipolar outflow, potentially linked to the interaction of the radio jet with the ambient gas \citep{Couto2020}. However, the galaxy-wide distribution and kinematics of the ionized gas remain insufficiently characterized. 

To further constrain the impact of AGN activity in 4C+29.30, we analyzed spatially resolved ionized gas kinematics and ionization diagnostics using optical IFS data from the Mapping Nearby Galaxies at Apache Point Observatory (MaNGA) survey \citep{Bundy2015}, covering a $32''$ diameter FoV extending to 1.75$R_e$ (effective radius) of the galaxy, together with new observations from the SITELLE \citep{Grandmont2012} instrument on the Canada-France-Hawaii Telescope (CFHT), which provides an $11' \times 11'$ FoV. Combining these data with radio imaging from the Very Large Array Sky Survey (VLASS; \citealt{Lacy2020}), matched to the MaNGA point-spread function ($\rm PSF \sim 2.5''$), we identify: (1) a localized redshifted region with enhanced velocity dispersion around the northern radio lobe, and (2) a galaxy-scale biconical outflow powered by the central AGN, whose axis is misaligned with the radio jet. These results suggest the coexistence of radiatively driven outflows and radio jets within the same system, while also demonstrating the impacts of the radio jets on the surrounding ionized gas. The paper is organized as follows. The data analysis is described in Section \ref{sec:data}. Results of kinematics and ionizations are presented in Section \ref{sec:results}. In Section 4, we discuss the implications of our findings for AGN feedback and its impacts on star formation. Finally, we summarize the conclusions in Section 5. 

\section{Data Analysis} \label{sec:data} 
\subsection{Multi-wavelength Imaging} \label{sec:images}

\begin{figure*}[ht!]
\centering
\includegraphics[width=\textwidth]{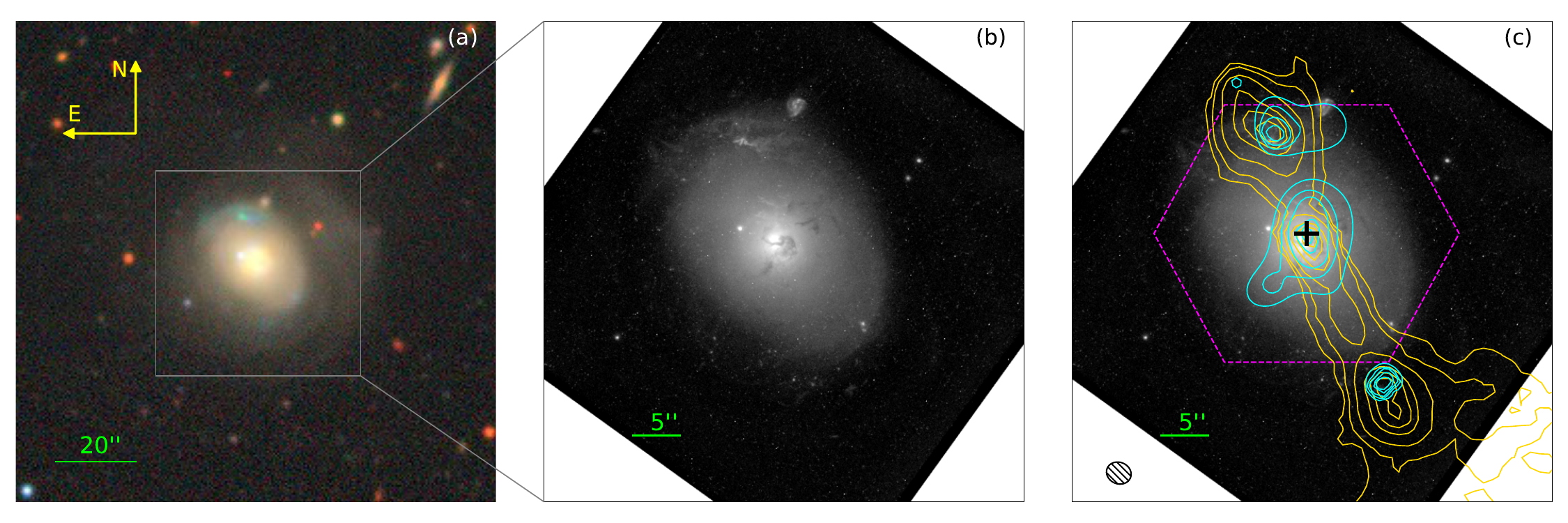}
\caption{Multi-wavelength observations of 4C+29.30.
(a) The $g$-, $r$-, and $z$-band composite image from the DESI Legacy Survey, where the central 45$''\times$45$''$ region is marked by a grey box. 
(b) The HST image centered at $\lambda=5852\text{\AA}$ with a bandwidth of $1873\text{\AA}$. 
(c) The HST image overlaid with the VLASS 3.0 GHz morphology shown as orange contours and the Chandra $0.5-2\rm\ keV$ X-ray emission shown as cyan contours. The VLASS beam size displayed in the lower-left corner is 2.84$''$$\times$2.28$''$ ($\sim$3.5 kpc $\times$ 2.8 kpc), and it has a position angle of 44$^\circ$ (east from north). The MaNGA bundle is marked by a magenta hexagon, and the black cross indicates the galaxy photometric center. 
\label{fig:morph}}
\end{figure*}

Figure \ref{fig:morph}(a) presents the $g$-, $r$-, and $z$-band composite image of 4C+29.30 from the Dark Energy Spectroscopic Instrument (DESI) Legacy Survey \citep{Dey2019}. It exhibits prominent merger-remnant features, including faint tidal streams, shells, and an asymmetric stellar halo. The broad-band imaging from the Hubble Space Telescope (HST) is shown in Figure \ref{fig:morph}(b), centered at $\lambda=5852\text{\AA}$ with a bandwidth of $1873\text{\AA}$ and a total exposure time of 2683 s \citep{Couto2020}. This deeper image, with its higher spatial resolution, resolves more detailed structures of the galaxy. In addition to ripple-like shells, a disturbed dust lane is clearly embedded in the bright central region. Figure \ref{fig:morph}(c) shows the same background image as in panel (b), where the superimposed radio morphology, including the jet and double lobes (orange), is spatially coincident with the Chandra $0.5-2\rm\ keV$ X-ray emission (cyan) from \cite{Siemiginowska2012}. 

The radio jet axis appears to be approximately parallel to the plane of the galactic disk, suggesting that the jet may propagate along and sweep across the disk surface before expanding to larger scales. This geometric configuration raises the question of whether the jet influences the ISM during this phase. To address this question, we analyze the optical IFS data of 4C+29.30 by combining the public MaNGA survey data (Section \ref{sec:manga}) with our follow-up SITELLE observations (Section \ref{sec:sitelle}).

\subsection{SDSS/MaNGA Data} \label{sec:manga}

The MaNGA survey \citep{Bundy2015,Drory2015,Wake2017} is one of the core projects of the fourth generation of SDSS (SDSS-IV; \citealt{Blanton2017}). It began in July 2014 and publicly released its full sample of $\sim$10,000 nearby galaxies in December 2021. MaNGA provides IFS observations using the Baryon Oscillation Spectroscopic Survey (BOSS) spectrographs \citep{Smee2013} on the 2.5-m Sloan Telescope \citep{Gunn2006}. Its ``hexabundle" FoV, determined by 17 fiber-bundle integral-field units, varies in diameter from 12$''$ (19 fibers) to 32$''$ (127 fibers). With a typical exposure time of 3 hours, the $r$-band signal-to-noise ratio (S/N) reaches $\rm 4-8\ \text{\AA}^{-1}$ per 2$''$ fiber at 1.5$R_e$. Given the spectral coverage (3,600$-$10,300\AA) and resolution ($R\sim 2000$) of MaNGA, together with the median redshift ($\langle z\rangle \sim$0.03) of the sample, we can analyze the key emission lines tracing the outflow gas (e.g. \oiii$\lambda\lambda$4959,5007, \oi$\lambda$6300), as well as the 4000\AA\ break index (D$_n$4000) as the tracer of luminosity-weighted stellar age. The stellar and ionized-gas observables are processed by the MaNGA data analysis pipeline (DAP; \citealt{Westfall2019}). This pipeline applies the penalized pixel-fitting (pPXF; \citealt{Cappellari2004}) method and a subset of stellar templates from the MaStar library \citep{Yan2019} to fit the stellar continuum, including absorption lines, in each spaxel, yielding measurements of the stellar kinematics. The parametric ionized-gas kinematics and flux are estimated by fitting a single Gaussian to each emission line after stellar continuum subtraction, where all emission lines are tied to have the same central velocity as the \ha\ emission.

4C+29.30 was observed within a 32$''$-diameter FoV, shown as a magenta dashed hexagon in Figure \ref{fig:morph}(c), which covers $\sim$1.75$R_e$ of the galaxy, where $R_e\sim9.14''$ as taken from the NASA-Sloan Atlas (NSA; \citealt{Blanton2011}). An angular size of 1$''$ corresponds to 1.245 kpc at the galaxy redshift of $z\sim0.0648$ in a cosmology with $H_0=70\rm\ km\ s^{-1}\ Mpc^{-1}$, $\Omega_{\rm m}=0.3$, and $\Omega_\Lambda=0.7$. The photometric position angle ($PA$) and axis ratio ($q$) are obtained from the NSA catalog, giving $PA=53^\circ$ (measured from north) and $q=0.8$. Based on the relation \citep{Hubble1926} between galaxy inclination ($i$) and $q$, ${\rm cos}^2 i = (q^2-q_0^2)/(1-q_0^2)$, we derive an inclination of $38^\circ$ for 4C+29.30 assuming $q_0=0.2$. 

\subsubsection{Velocity Channel Map} \label{sec:channel}

\begin{figure*}[ht!]
\centering
\includegraphics[width=\textwidth]{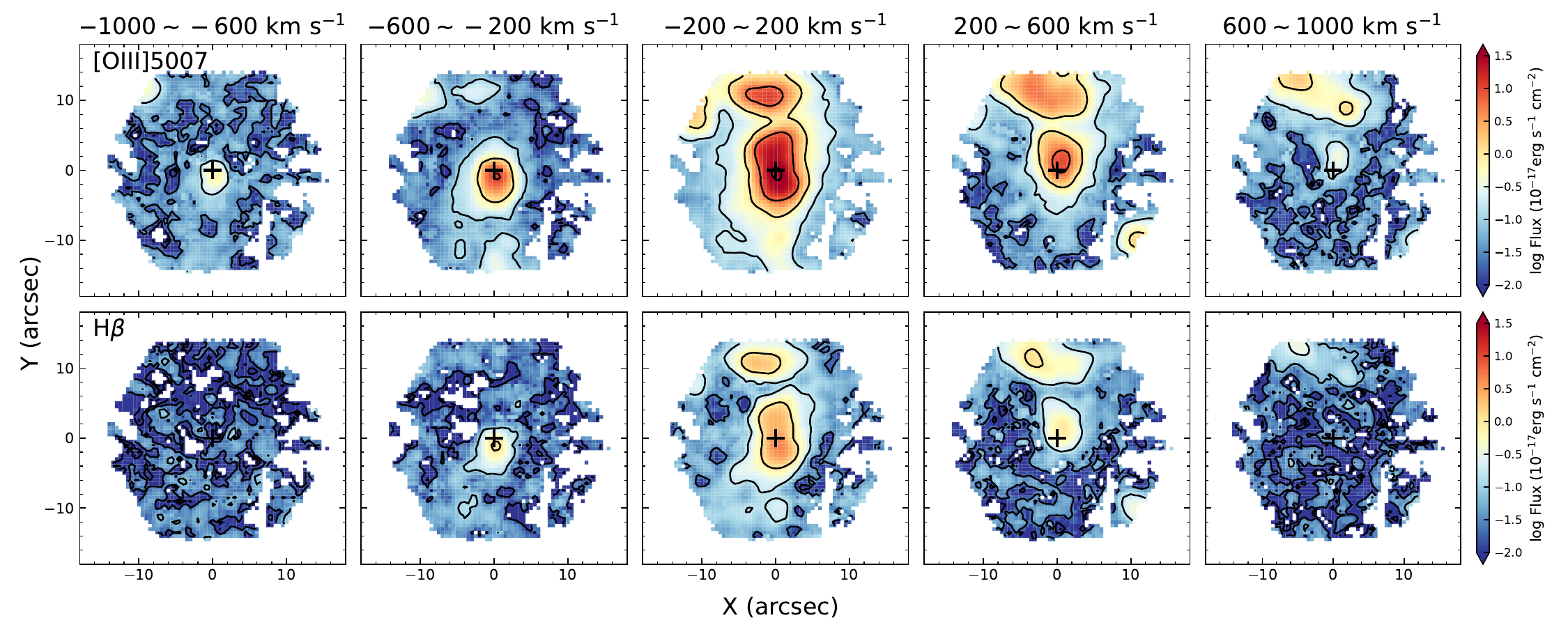}
\caption{Velocity channel maps of \oiii$\lambda$5007 (top) and \hb\ (bottom). For each emission line, the velocity ranges from $-1000\rm\ km\ s^{-1}$ to $1000\rm\ km\ s^{-1}$, divided into bins of $400\rm\ km\ s^{-1}$. Superimposed black contours highlight regions of enhanced flux in each channel.
\label{fig:channel-O3Hb}}
\end{figure*}

We analyze non-parametric velocity channel maps for emission lines that are not significantly contaminated by nearby transitions (e.g., \ha\ is blended with \nii$\lambda\lambda$6548,6585). The velocity channel maps of \oiii$\lambda$5007 and \hb\ are shown in Figure \ref{fig:channel-O3Hb}. For each emission line, the channel maps span velocities from $-1000\ \rm km\ s^{-1}$ to $1000\ \rm km\ s^{-1}$ with bins of $400\ \rm km\ s^{-1}$. Both \oiii$\lambda$5007 and \hb\ exhibit enhanced flux in the northern region, and this region is predominantly redshifted, with velocities extending beyond $600\ \rm km\ s^{-1}$. These features motivate a parametric emission-line fitting approach using multiple Gaussian components to better characterize the ionized gas kinematics across the galaxy.

\subsubsection{Parametric Emission-line Fitting} \label{sec:linefit}

We first perform an initial fit over the spectral cube using a single Gaussian component for each emission line after stellar continuum subtraction, followed by a second fit using two Gaussian components. The parametric fitting procedure is carried out using the MPFIT algorithm, which minimizes the reduced chi-square ($\chi^2$) via the Levenberg-Marquardt least-squares method. We focus on the emission lines \ha, \nii$\lambda\lambda$6548,6585, \sii$\lambda\lambda$6718,6732, \hb, and \oiii$\lambda\lambda$4959,5007, which are effective for classifying the ionization mechanism and tracing the gas-phase metallicity. Since the \oiii$\lambda\lambda$4959,5007 line profiles are not typically consistent with those of the other emission lines (e.g., \citealt{Reines2013,Cao2025}), we fit the \oiii\ lines independently, requiring \oiii$\lambda$4959 and \oiii$\lambda$5007 to share the same velocity centroid and velocity dispersion. The \nii$\lambda\lambda$6548,6585, \sii$\lambda\lambda$6718,6732, and \hb\ lines are tied to have the same velocity centroid as \ha, while their velocity dispersions are allowed to vary within 0.75$-$1.25 times that of \ha.

\begin{figure*}[ht!]
\plotone{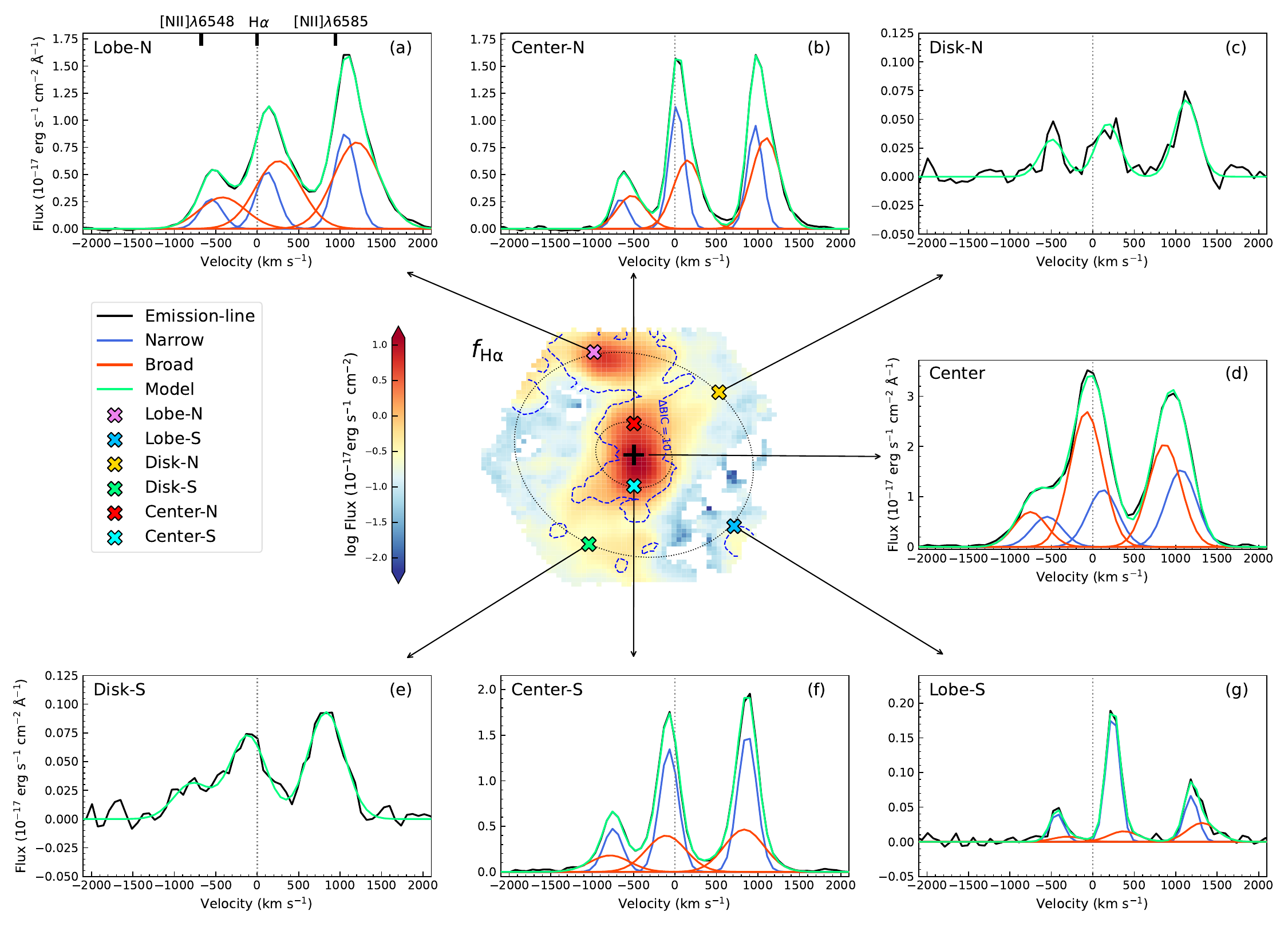}
\caption{\ha\ emission-line spectral fitting. The central map is the non-parametric \ha\ flux derived by DAP. The contour representing $\rm \Delta BIC=10$ for the \ha+\nii\ wavelength range is marked by a blue dashed curve. The larger black dotted ellipse, centered on the galaxy photometric center (black cross), covers the flux peak (purple cross, Lobe-N) of the northern radio lobe. Its geometric parameters are defined according to the optical morphology of the galaxy. A spaxel symmetric to Lobe-N is located at the boundary of the southern radio lobe (blue cross, Lobe-S). Two representative spaxels from the northern and southern galactic disk are marked by yellow (Disk-N) and green (Disk-S) crosses. A second black dotted ellipse with a smaller diameter is defined with the same geometric parameters as the larger one. Two mutually symmetric spaxels on the northern and southern sides of the smaller ellipse are marked by red (Center-N) and cyan (Center-S) crosses. Panels (a-g) present the \ha+\nii\ spectral fitting results for these selected spaxels marked on the flux map, with the label of each spaxel shown at the top-left corner of the corresponding panel. The rest-frame positions of the \ha\ and \nii$\lambda$6548,6585 emission lines are marked at the top of panel (a) in velocity space. The velocity axis is defined relative to the rest-frame wavelength of the \ha\ emission line, and the velocity measurements are derived from the \ha\ emission line. In each panel, the black line shows the observed emission-line spectrum; the blue (narrower) and red (broader) lines represent the two Gaussian components. Their combination is shown in green, indicating the best-fitting model. The emission-line spectra at Disk-N (c) and Disk-S (e) are fitted with a single Gaussian component.
\label{fig:fit-ha}}
\end{figure*}

\begin{figure*}[ht!]
\plotone{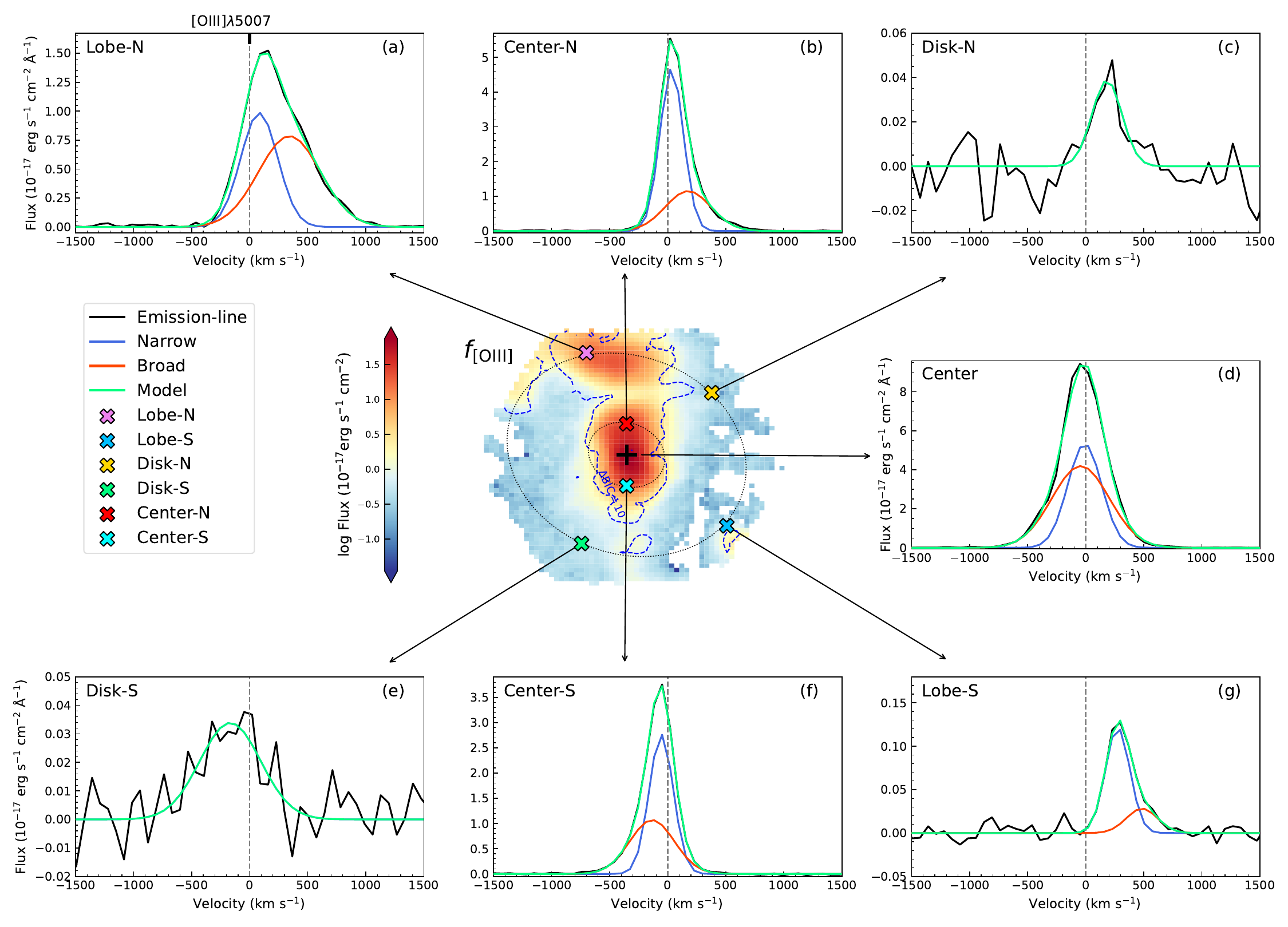}
\caption{\oiii\ emission-line spectral fitting. The non-parametric \oiii\ flux from DAP is presented in the central panel, in which the blue dashed line represents the contour of $\rm \Delta BIC=10$ for the \oiii$\lambda$5007 line. Panels (a-g) show the \oiii\ emission-line spectral fitting results for the selected spaxels marked on the flux map. The configuration of each selected spaxel and the corresponding spectral panel is the same as in Figure \ref{fig:fit-ha}. 
\label{fig:fit-o3}}
\end{figure*}

To quantify the goodness of the single- and double-Gaussian models, we compute the Bayesian Information Criterion (BIC; \citealt{ReichardtChu2022}) for each fit. The difference in BIC between the single ($\rm BIC_1$) and double ($\rm BIC_2$) Gaussian models, defined as $\Delta \rm BIC = BIC_1 - BIC_2$, serves as a criterion to evaluate whether the additional component significantly improves the fit. The \ha+\nii\ wavelength range of $6530\ \text{\AA} < \lambda < 6620\ \text{\AA}$ is used to calculate $\Delta \rm BIC$, while for \oiii$\lambda$5007 we adopt $4990\ \text{\AA} < \lambda < 5030\ \text{\AA}$. Spaxels with $\Delta \rm BIC > 10$ are selected for double-Gaussian modeling \citep{Swinbank2019,Avery2021}. Under this criterion, more than 95\% of the selected emission lines show at least a 50\% improvement in $\chi^2$ relative to the single-Gaussian model (i.e., a $\chi^2$ ratio between the single- and double-Gaussian fittings is larger than 1.5). The $\Delta \rm BIC = 10$ contour is overlaid on the non-parametric \ha\ and \oiii\ flux maps from the DAP (Figures \ref{fig:fit-ha} and \ref{fig:fit-o3}). The flux-enhanced northern region, dominated by redshifted velocity components, lies largely within this contour, and the corresponding spaxels are therefore modeled with two Gaussian components.

Figure \ref{fig:fit-ha} presents representative \ha+\nii\ fitting examples at different spatial locations. In the central \ha\ flux map, the outer ellipse (black dotted ellipse), centered on the photometric center of the galaxy, has geometric parameters ($PA$ and $q$) set according to the galaxy morphology and encompasses the flux peak of the northern radio lobe (purple cross; Lobe-N). A spaxel symmetric to Lobe-N along this ellipse is located near the southern radio lobe (blue cross; Lobe-S). Two additional spaxels in the northern and southern disk regions are labeled Disk-N (yellow cross) and Disk-S (green cross), respectively. An inner ellipse with the same geometric parameters is defined closer to the galaxy center. Two symmetric spaxels along this ellipse are labeled Center-N (red cross) and Center-S (cyan cross). Selecting spaxels at similar galactocentric radii along the same ellipse minimizes the impact of radial gradients and structural differences within the galaxy.

In the Lobe-N example (Figure \ref{fig:fit-ha}a), the black line represents the observed spectrum; the blue and red curves correspond to the two Gaussian components, with the blue curve denoting the narrower component. Their combination (green curve) represents the best-fit model. The same fitting configuration is adopted for all selected regions (panels a-g).

For Lobe-N and Lobe-S (panels a and g), both components are redshifted. Center-N (panel b) shows two redshifted components, while the symmetric Center-S (panel f) exhibits two blueshifted components. In the central spaxel (panel d), one component is blueshifted and the other redshifted, both relative to the systemic velocity. In the disk regions (panels c and e), a single Gaussian component suffices, with Disk-N and Disk-S showing redshifted and blueshifted velocities, respectively. Overall, the northern side of the galaxy is predominantly redshifted relative to the southern side. The regions around the radio lobes display more complex kinematics, including broader line widths and larger velocity offsets. Similar trends are observed in the \oiii$\lambda$5007 fitting results (Figure \ref{fig:fit-o3}).

From the parametric fitting of emission lines with $\rm S/N > 3$, the narrow components in the double-Gaussian models have velocity dispersions of $\rm 50-150\ \rm km\ s^{-1}$ and show good continuity with the disk rotation outside the jet-dominated region. We define these narrow components, together with the single-Gaussian fits, as Component 1 (C1), which likely traces a rotating gaseous disk. The broader components are defined as Component 2 (C2), with velocity dispersions between 150 and 500 $\rm km\ s^{-1}$.

\subsection{CFHT/SITELLE Observations} \label{sec:sitelle}

To complement the MaNGA observations, we used the SITELLE \citep{Grandmont2012} IFS instrument at CFHT. SITELLE has an $11' \times 11'$ FoV, which encompasses the entire region of radio emission. 4C+29.30 was observed on February 3, 2025, under $\sim1''$ seeing conditions, with a total exposure time of 6.5 hours. SITELLE is an imaging Fourier transform spectrometer (IFTS) designed for CFHT, providing IFS capabilities over the visible wavelength range of 350$-$900 nm, with a variable spectral resolution of $R = 2-10,000$ \citep{Drissen2019}. We employed the C3 filter (511$-$556 nm) to cover the key \hb\ and \oiii$\lambda\lambda$4959,5007 emission lines. The resulting spectral resolution ($R \sim 1945$) is comparable to that of the MaNGA observations.

The calibrated data cubes were retrieved from the Canadian Astronomy Data Centre (CADC).\footnote{\url{https://www.cadc-ccda.hia-iha.nrc-cnrc.gc.ca/en/}} To improve the S/N of extended emission in the outskirts of the galaxy, the spectral cube was spatially rebinned by a factor of 2, corresponding to a pixel scale of $0.64''$. We used the Python module ORCS \footnote{\url{https://github.com/thomasorb/orcs}} \citep{Martin2015} to fit the stellar continuum and emission lines in the SITELLE spectral cubes. Given the relatively flat continuum shape and the absence of prominent absorption features in the observed spectral range, a constant function is sufficient to model the stellar continuum. Emission lines are fitted using a ``sincgauss" profile, which combines the intrinsic instrumental line shape (sinc function) with a Gaussian component. As a result, our analysis focuses on relative comparisons of emission-line velocity dispersions within the galaxy. 

Using the same parametric fitting strategy adopted for the MaNGA data, we perform both single- and double-component fits for the SITELLE emission lines. The narrower component in the double-component model, together with the single-component fits, is defined as SITELLE Component 1 (SC1), while the broader component in the double-component model is defined as SITELLE Component 2 (SC2).

\section{Results} \label{sec:results}

In this section, we present the spatially resolved kinematics and ionization properties of the galaxy. Both non-parametric and parametric emission-line diagnostics are used to characterize the ionized gas.

\subsection{Stellar and Ionized Gas Kinematics} \label{sec:kinematics}
\subsubsection{Gas-star Misalignment} \label{sec:misalign}

\begin{figure*}[ht!]
\centering
\includegraphics[width=\textwidth]{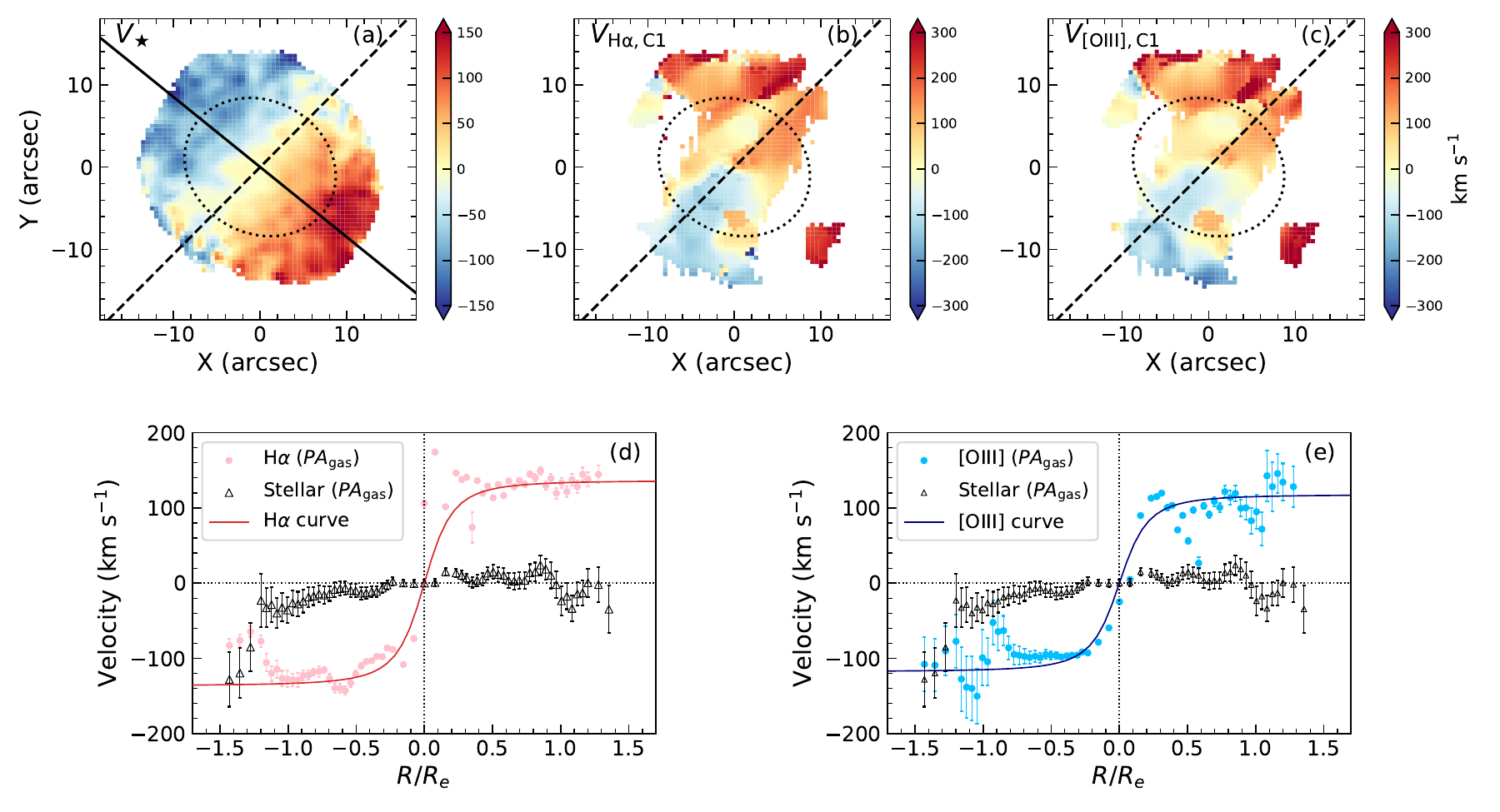}
\caption{Gas-star misalignment. 
(a) Stellar velocity field from DAP. 
(b,c) \ha\ ($V_{\rm H\alpha,C1}$) and \oiii\ ($V_{\rm [OIII],C1}$) velocity fields of the decomposed Component-1, which combines the single-Gaussian component and the narrower component from double-Gaussian fitting. The solid and dashed lines in each velocity field represent the kinematic position angle ({\it PA}) of the stellar and decomposed gaseous components, respectively. The {\it PA} is calculated using spaxels within 1$R_e$ (dotted ellipse) of the galaxy. 
(d) Velocities as a function of radius along the gaseous position angle ({\it PA}$_{\rm gas}$) for stars (black trangles) and \ha\ (red dots), where each error bar shows the 1$\sigma$ uncertainty. The best-fitted velocity curve of \ha\ velocities is shown by a red solid line. 
(e) The same configuration as in panel (d), but the blue dots represent the \oiii\ velocities, and the corresponding best-fitting model is shown by a blue solid curve.
\label{fig:misalign}}
\end{figure*}

The stellar kinematics of 4C+29.30 generally exhibit a regular rotation pattern as probed by MaNGA, spanning velocities from $-150\ \rm km\ s^{-1}$ to $150\ \rm km\ s^{-1}$ (Figure \ref{fig:misalign}a). The gaseous kinematics are decomposed into two components through our emission-line fitting procedure (Section \ref{sec:linefit}). The first kinematic component (C1), defined as the combination of the narrower component in the double-Gaussian model and the single-Gaussian fits, likely traces a rotating gaseous disk. Figure \ref{fig:misalign}(b,c) present the C1 velocity fields of \ha\ and \oiii. The second kinematic component (C2), corresponding to the broader Gaussian component, is discussed in Section \ref{sec:accelerate}.

Using the stellar and C1 gas velocity fields, we measure their kinematic position angles ($PA$) within $R_e$ (dotted ellipse in Figure \ref{fig:misalign}a-c, corrected for inclination) following the method of \cite{Krajnovic2006}. The kinematic $PA$ is defined as the counter-clockwise angle from north to the line that bisects the velocity field on the approaching side. We derive $PA_\star \sim 50^\circ$ for the stellar component and $PA_{\rm gas} \sim 137^\circ$ for both the \ha\ and \oiii\ ionized gas. The corresponding major axes are shown as solid and dashed lines in Figure \ref{fig:misalign}(a-c). The difference between $PA_\star$ and $PA_{\rm gas}$ is $\Delta PA \sim 87^\circ$, which significantly exceeds the commonly adopted misalignment threshold of $30^\circ$ \citep{Chen2016}. This indicates that the kinematic major axis of the ionized gas is strongly misaligned with respect to the stellar component.

Figure \ref{fig:misalign}(d,e) show the stellar and gaseous velocities as a function of radius measured along $PA_{\rm gas}$. The \ha\ (pink dots) and \oiii\ (blue dots) velocities span approximately $-150\ \rm km\ s^{-1}$ to $150\ \rm km\ s^{-1}$, whereas the stellar velocities (black triangles) remain close to $0\ \rm km\ s^{-1}$ along this axis. We fit the gas velocities using a rotation curve model following \cite{Bertola1991}, shown as solid lines for \ha\ (red) and \oiii\ (blue). The good agreement between the model and the observed gas velocities supports the presence of a rotating gaseous disk whose kinematic axis is nevertheless significantly misaligned with the stellar disk. In addition, deviations from pure circular rotation are primarily found away from the kinematic major axis, particularly toward the northern region, and may be associated with jet-ISM interactions or residual effects of the past gas-rich merger.

\subsubsection{Gas Acceleration and Outflow} \label{sec:accelerate}

\begin{figure*}[ht!]
\plotone{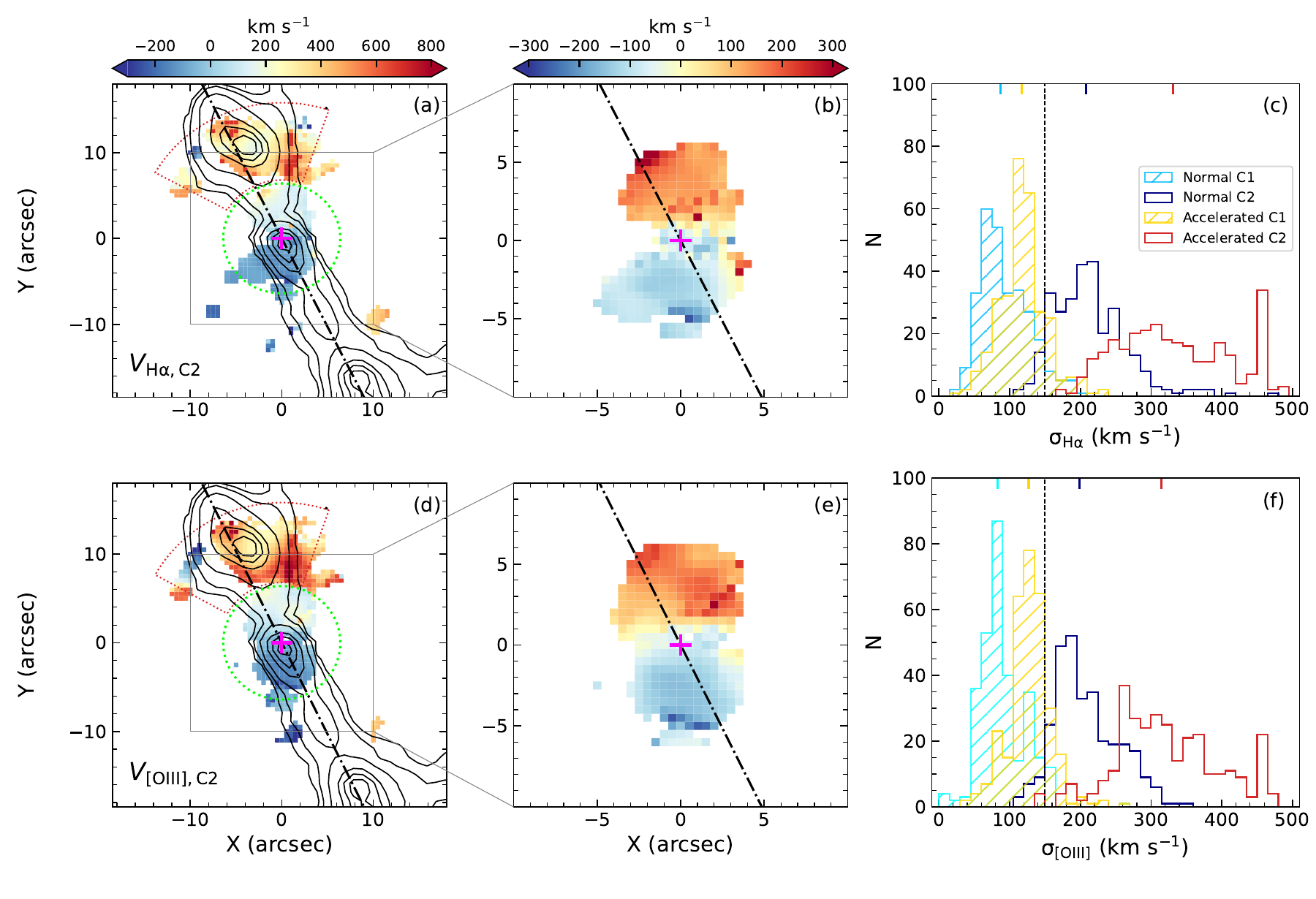}
\caption{Kinematics of the decomposed broader gas component. 
(a) \ha\ velocity field for the decomposed Component-2 ($V_{\rm H\alpha,C2}$), corresponding to the broader component identified from double-Gaussian fitting. The VLASS 3.0 GHz radio emission is shown as black contours. Its major axis, oriented at $\sim 26^\circ$ east of north, is marked by a dot-dashed line. The northern radio lobe coincides with a region with enhanced velocities, marked by a red dotted sector. The central region without strong redshifted velocities is indicated by a green dotted circle with a radius of 0.7$R_e$ ($\sim$8 kpc). 
(b) Zoomed-in velocity field of the central labeled region, where the colorbar range ($-300\ \rm km\ s^{-1}$ to $300\ \rm km\ s^{-1}$) is smaller than that in panel (a) ($-300\ \rm km\ s^{-1}$ to $800\ \rm km\ s^{-1}$). 
(c) Velocity dispersion distributions. The yellow (Accelerated C1) and red (Accelerated C2) histograms represent the C1 and C2 components in the accelerated region around the northern radio lobe (red dotted sector in panel a), respectively. The cyan (Normal C1) and blue (Normal C2) histograms correspond to the non-accelerated region, defined by spaxels within the green circle shown in panel (a). The black dotted line indicates a velocity dispersion of $150\ \rm km\ s^{-1}$. 
Panels (d-f) present the kinematics of the decomposed broader component for \oiii.
\label{fig:accelerate}}
\end{figure*}

The velocity field of the second kinematic component (C2) traced by \ha\ is shown in Figure \ref{fig:accelerate}(a), with black contours indicating the VLASS 3.0 GHz radio emission. The radio major axis (dot-dashed line) is oriented at a position angle of $\sim26^\circ$ (east of north; \citealt{Jamrozy2007}). A prominent region of enhanced velocities within a red dotted sector is observed around the northern radio lobe. In contrast, the region marked by the green dotted circle does not exhibit such high-velocity features.
To better resolve the inner kinematics, we focus on the central $10'' \times 10''$ region (Figure \ref{fig:accelerate}b). The zoomed-in panel adopts a narrower color scale ($-300\ \rm km\ s^{-1}$ to $300\ \rm km\ s^{-1}$) than that used in panel (a) ($-300\ \rm km\ s^{-1}$ to $800\ \rm km\ s^{-1}$). Within the green-circled region (corresponding to $0.7R_e \approx 8$ kpc), the velocity field exhibits almost symmetric kinematics, with redshifted velocities to the north and blueshifted velocities to the south relative to the galaxy center. Figures \ref{fig:accelerate}(d,e) present the same analysis for the \oiii\ C2 velocity field, showing similar spatial trends.

We further compare the velocity dispersion distributions of C1 and C2 in the accelerated region within the red dotted sector and the non-accelerated region marked by the green circle in Figure \ref{fig:accelerate}(a). In Figure \ref{fig:accelerate}(c), the yellow (Accelerated C1) and red (Accelerated C2) histograms represent the C1 and C2 components in the velocity-enhanced region, while the cyan (Normal C1) and blue (Normal C2) histograms correspond to the non-accelerated region. Median velocity dispersions are indicated by short solid lines. Figure \ref{fig:accelerate}(f) shows the corresponding \oiii\ distributions. In both regions, C2 components are broader than C1 by construction of the kinematic decomposition. Notably, C2 velocity dispersions generally exceed $150\ \rm km\ s^{-1}$, a regime commonly associated with shock-ionized gas \citep{Kewley2019}. Moreover, only the velocity-enhanced region exhibits C2 dispersions extending beyond $400\ \rm km\ s^{-1}$. These results suggest localized gas acceleration in the vicinity of the northern radio lobe, consistent with jet-ISM interaction. Within the central $0.7R_e$ region, the velocity structure is broadly consistent with a biconical ionized gas outflow whose axis is offset by $\sim26^\circ$ from the radio jet.

\begin{figure*}[ht!]
\centering
\includegraphics[width=\textwidth]{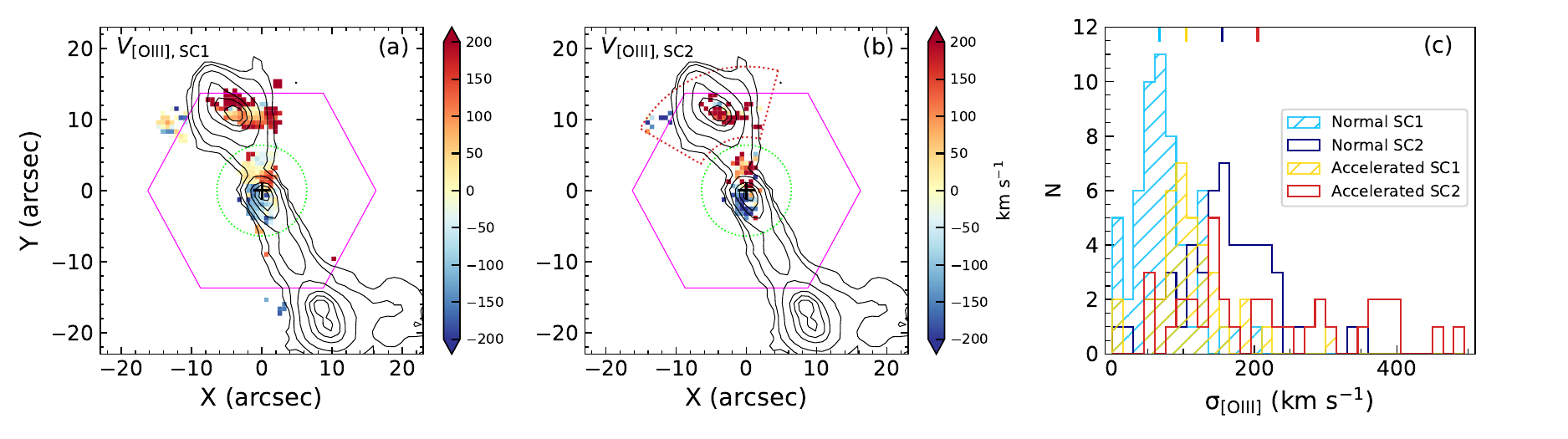}
\caption{Decomposed \oiii\ kinematics based on SITELLE observations. 
(a) The velocity field of the fitted Component-1 derived from the SITELLE spectral cube (SC1). 
(b) The velocity field of Component-2 derived from fitting the SITELLE data (SC2). 
In each velocity field, the overlaid magenta hexagon represents the MaNGA bundle; black contours show the $\rm VLASS\ 3.0\ GHz$ emission. The green dotted circle is the same as in Figure \ref{fig:accelerate}. The red dotted sector marks the accelerated region around the northern radio lobe. 
(c) Velocity dispersion distributions for the two decomposed components in the accelerated (yellow for SC1, red for SC2) and non-accelerated (cyan for SC1, blue for SC2) regions.
\label{fig:sitelle}}
\end{figure*}

Figure \ref{fig:sitelle}(a,b) show the \oiii\ velocity fields of the two SITELLE components (SC1 and SC2). Within the MaNGA footprint, the SITELLE kinematics closely resemble those derived from MaNGA. In the SC2 map, enhanced velocities are again observed around the northern radio lobe (red dotted sector), and a biconical-like structure is evident in the central $0.7R_e$ region (green dotted circle). The \oiii\ velocity dispersion distribution (Figure \ref{fig:sitelle}c) shows that SC2 in the accelerated region exhibits systematically higher dispersions relative to other regions of the galaxy, further supporting the jet-ISM interaction scenario indicated by the MaNGA data.

\subsection{Ionization Mechanisms} \label{sec:ionize}

To constrain the origin of the observed gas acceleration and possible outflow, we use emission-line diagnostic diagrams to diagnose the spatially resolved ionization sources across the galaxy. We first analyze the non-parametric integrated emission-line flux measured within the velocity range of $-600\ \rm km\ s^{-1}$ to $600\ \rm km\ s^{-1}$. This velocity interval is chosen to encompass both the regular rotational component and the irregular (accelerated or outflowing) component, while minimizing contamination from adjacent emission lines. Given that the narrower and broader kinematic components are decomposed across the full MaNGA FoV (Section \ref{sec:linefit}), we further examine the ionization properties of these two components separately, providing complementary constraints to the non-parametric analysis.

\subsubsection{Emission-line Ratios} \label{sec:lineratio}

\begin{figure*}[ht!]
\plotone{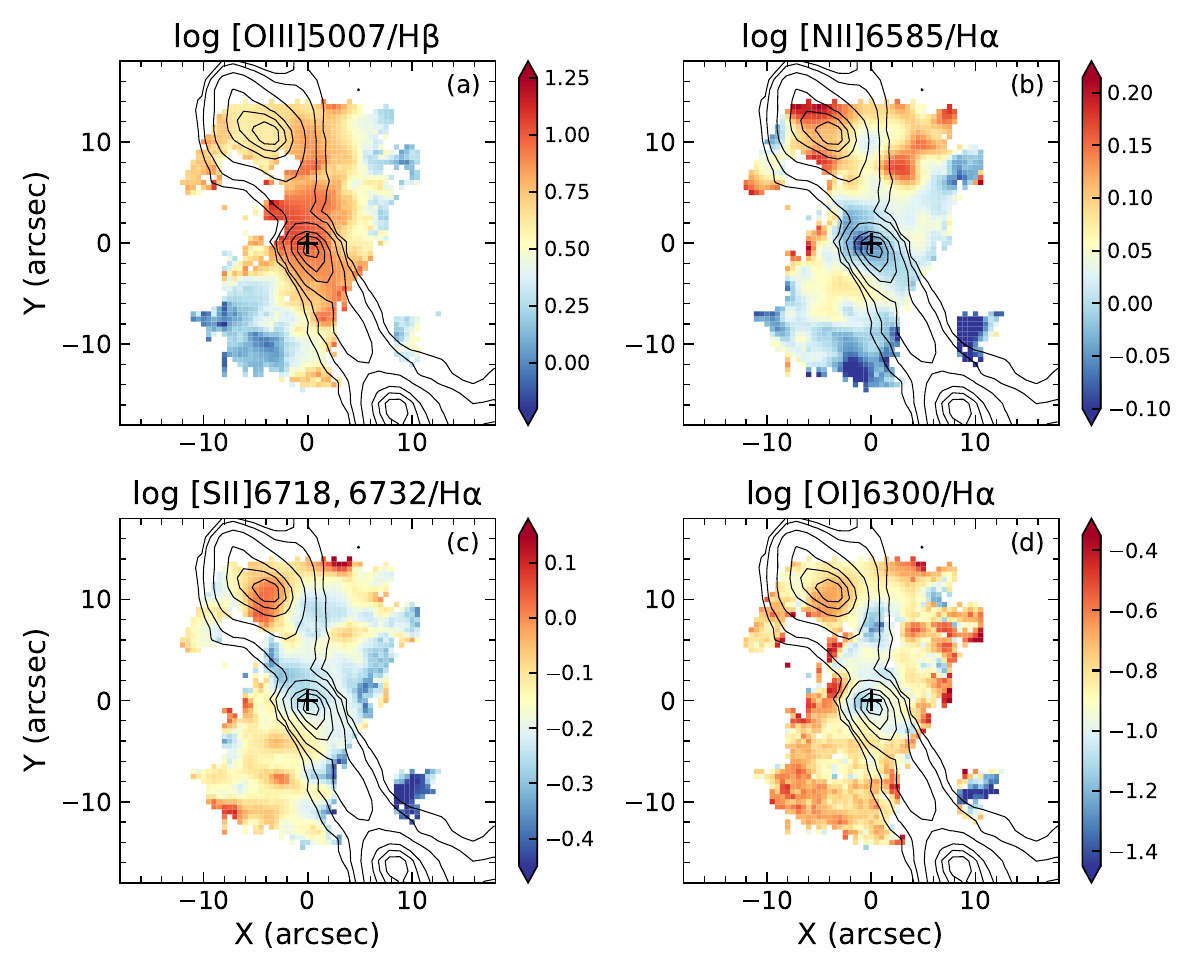}
    \caption{Emission-line ratio maps.
    (a) \oiii/\hb.
    (b) \nii/\ha.
    (c) \sii/\ha.
    (d) \oi/\ha.
    The logarithms of the line-flux ratios are color-coded according to the color bar. 
    Black contours show the VLASS 3.0 GHz emission.
    } 
    \label{fig:lineRatio}
\end{figure*}

Figure \ref{fig:lineRatio} presents the spatially resolved emission-line ratio maps of \oiii/\hb, \nii/\ha, \sii/\ha, and \oi/\ha, derived using non-parametric emission-line flux. The black contours indicate the 3.0 GHz radio emission. The central region exhibits elevated \oiii/\hb\ together with comparatively lower \nii/\ha, \sii/\ha, and \oi/\ha, suggestive of a high ionization parameter consistent with Seyfert-like AGN photoionization. At larger radii, \oiii/\hb\ decreases, while the low-ionization line ratios (\nii/\ha, \sii/\ha, and \oi/\ha) increase relative to the center, likely consistent with the ionization conditions of low-ionization nuclear emission-line region (LINER). Notably, the region surrounding the northern radio lobe shows intermediate \oiii/\hb\ values, lower than in the galaxy center but higher than in the surrounding disk. Furthermore, its low-ionization line ratios are significantly elevated compared with all other regions of the galaxy. This line-ratio pattern, in combination with its spatial association with the radio emission, suggests the presence of a composite ionization mechanism that may involve both AGN photoionization and shock excitation related to the radio jet.

\subsubsection{Non-parametric Diagnostics} \label{sec:nonparam-ionize}

\begin{figure*}[ht!]
\plotone{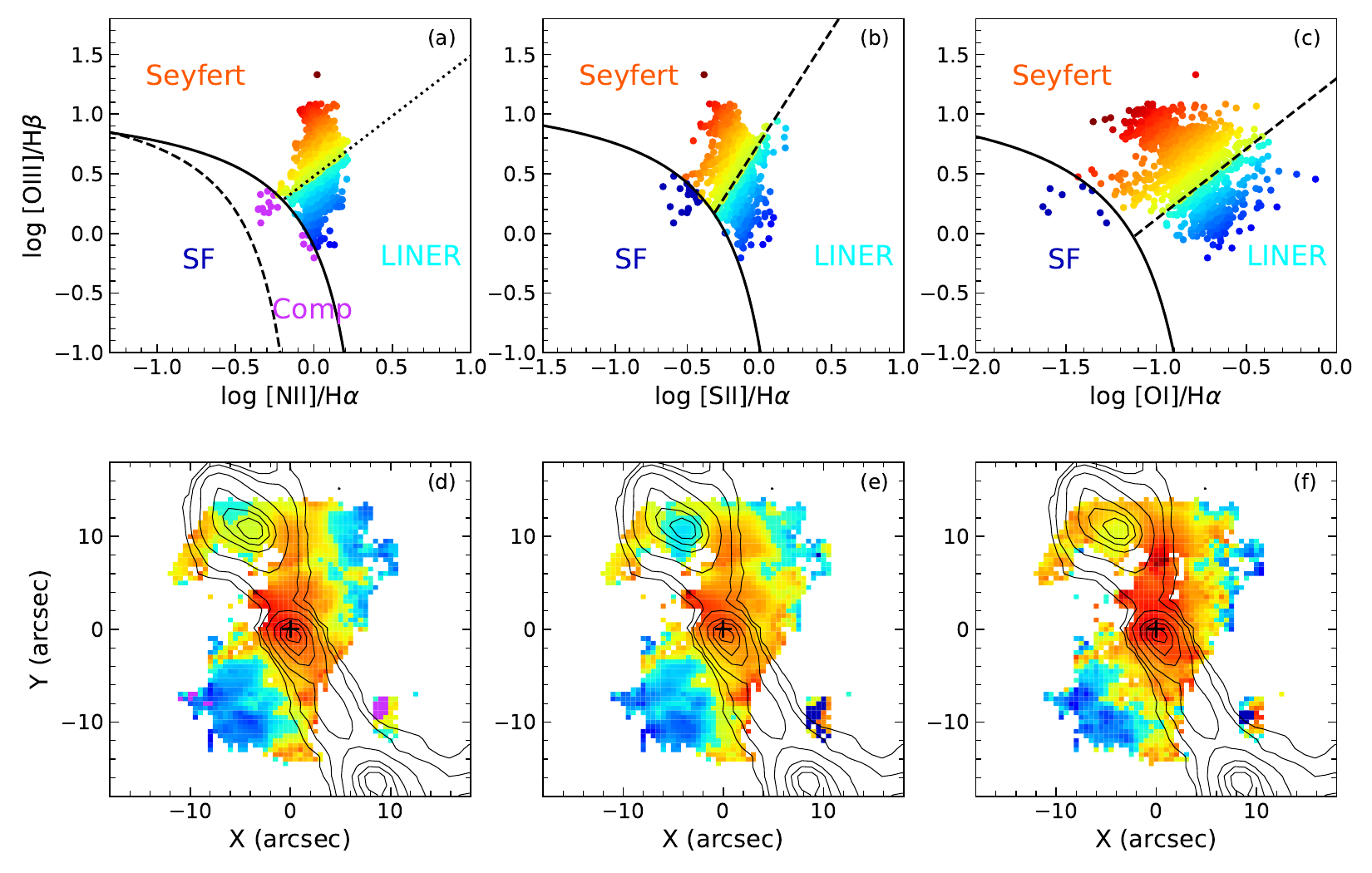} 
    \caption{Emission-line ratio diagnostic diagrams. 
    (a) \nii/\ha\ versus \oiii/\hb\ diagram. The dashed and solid curves \citep{Kauffmann2003,Kewley2001} represent the demarcation lines separating star-forming (SF), composite (Comp, magenta), and AGN regions. AGN-like spaxels are further divided into Seyfert and LINER by the dotted line \citep{CidFernandes2010}, where the spaxels are color-coded according to their distance from the dotted demarcation line. A color gradient from yellow to red is used for Seyfert spaxels and from green to blue for LINER spaxels, indicating increasing distance from the demarcation line. 
    (b) \sii/\ha\ versus \oiii/\hb\ diagram. The solid and dashed curves \citep{Kewley2001,Kewley2006} separate SF (blue), Seyfert, and LINER regions. 
    (c) \oi/\ha\ versus \oiii/\hb\ diagram. The classification scheme is the same as in panel (b), but applied to the \oi/\ha\ line ratios. 
    The Seyfert and LINER spaxels in the \sii\ and \oi\ diagrams adopt the same color-coding scheme as in the \nii\ diagram.
    (d-f) Spatially resolved classification maps with color coding consistent with the corresponding diagnostic diagrams shown in panels (a-c). Black contours show the VLASS 3.0 GHz emission.
    \label{fig:classify-colorful}}
\end{figure*}

Using the derived emission-line ratios, we construct the \nii, \sii, and \oi\ diagnostic diagrams \citep{Baldwin1981,Veilleux1987}, in which \oiii/\hb\ is plotted as a function of \nii/\ha, \sii/\ha, and \oi/\ha, respectively. The \nii/\ha$-$\oiii/\hb\ diagram is shown in Figure \ref{fig:classify-colorful}(a) for spaxels with $\rm S/N>3$ in all four lines. The black solid curve represents the theoretical maximum starburst line \citep{Kewley2001}, above which spaxels are classified as AGN-like. The black dashed curve indicates the empirical demarcation derived by \cite{Kauffmann2003}, separating purely star-forming regions from composite ones. The region between these two curves is classified as composite (Comp, magenta), indicating contributions from both AGN and star formation. The black dotted line \citep{CidFernandes2010} further separates AGN-like spaxels into Seyfert and LINER regions. Figures \ref{fig:classify-colorful}(b,c) show the \sii/\ha$-$\oiii/\hb\ and \oi/\ha$-$\oiii/\hb\ diagrams. The solid \citep{Kewley2001} and dashed \citep{Kewley2006} curves divide the spaxels into Seyfert, LINER, and star-forming (SF, blue) regions.

To better quantify the ionization conditions of AGN-like regions, we measure the perpendicular distance of each AGN spaxel from the Seyfert-LINER demarcation line in the three diagnostic diagrams, extending a method recently explored in the literature \citep{Krol2025}. In Figure \ref{fig:classify-colorful}(a,b\&c), the color scale from yellow to red (or green to blue) represents the increasing distance from the demarcation line within the Seyfert (or LINER) regime. The corresponding spatially resolved classification maps based on \nii, \sii, and \oi\ diagnostics are presented in Figure \ref{fig:classify-colorful}(d,e\&f), with color codes consistent with the diagnostic diagrams. The radio morphology is overlaid as black contours in each panel.
The \nii, \sii, and \oi\ diagnostic diagrams present consistent results in the northern radio lobe, where the region is dominated by spaxels located close to the Seyfert-LINER boundary, suggesting that both ionization mechanisms may contribute in this region. Combined with the enhanced velocities and elevated velocity dispersions around the radio lobe (Figure \ref{fig:accelerate}), these results are consistent with a scenario in which jet-driven shocks disturb the surrounding ISM and significantly influence the local ionization conditions.

\subsubsection{Parametric Diagnostics} \label{sec:param-ionize}

Based on the parametric emission-line fitting of the MaNGA data cube, two kinematic components are resolved from the emission lines. One component is likely associated with a rotating disk (C1), while the other traces non-circular motions (C2). We analyze the ionization properties of these two components separately in order to distinguish emission associated with the rotating disk from that related to outflows or jet interactions.

\begin{figure*}[ht!]
\plotone{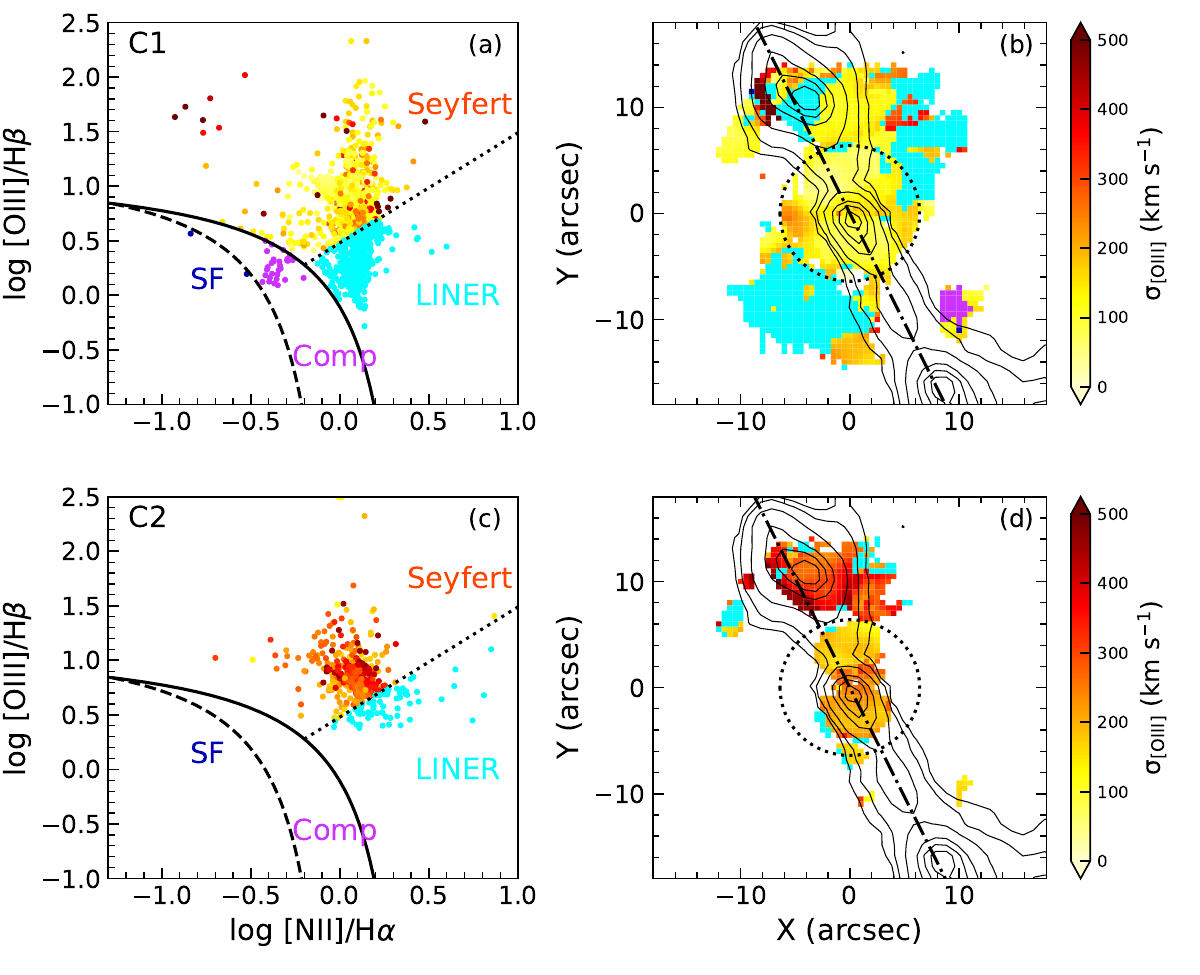} 
    \caption{Parametric \nii/\ha\ versus \oiii/\hb\ diagnostic diagrams. 
    (a) Ionization classifications for Component-1 (C1), which consists of the single Gaussian component and the narrower component from double-Gaussian fitting. The dashed and solid \citep{Kauffmann2003,Kewley2001} curves represent the demarcation lines separating the star-forming (SF, blue), composite (Comp, magenta), and AGN regions. The dotted line \citep{Kewley2006} further separates AGN-like spaxels into Seyfert and LINER (cyan). Seyfert spaxels are color-coded according to their \oiii\ velocity dispersion, with a color gradient from yellow to red indicating increasing velocity dispersion. 
    (b) The corresponding spatially resolved diagnostic map with color coding consistent with the corresponding diagnostic diagram shown in panel (a). Black contours indicate the VLASS 3.0 GHz radio emission and its major-axis is marked by a dot-dashed line. The dotted circle indicates the central 0.7$R_e$ region of the galaxy.
    Panels (c,d) show the same configuration as in panels (a,b), but for Component-2 (C2), corresponding to the broader component identified from double-Gaussian fitting.
    \label{fig:bpt-n2}}
\end{figure*}

Figure \ref{fig:bpt-n2}(a) presents the \nii/\ha$-$\oiii/\hb\ diagnostic diagram for the C1 component, including spaxels with $\rm S/N>3$ in all four emission lines. Star-forming, composite, and LINER spaxels are shown in blue, magenta, and cyan, respectively. Seyfert spaxels are color-coded according to their corresponding \oiii\ velocity dispersions. The spatially resolved classification map is shown in Figure \ref{fig:bpt-n2}(b). The central region within 0.7$R_e$ (black dotted circle) is dominated by Seyfert-like ionization, whereas the outer disk is primarily classified as LINER. Around the northern radio lobe, LINER spaxels become more prominent. Combined with the irregular gas kinematics in this region (Figure \ref{fig:misalign}b,c), this behavior is consistent with additional shock excitation likely related to jet-ISM interaction. A composite region, indicating contributions from both star formation and AGN activity, is located near the boundary of the southern radio lobe.

The \nii/\ha$-$\oiii/\hb\ diagram for the C2 component and its corresponding spatial classification map are presented in Figure \ref{fig:bpt-n2}(c,d). In contrast to C1, the C2 component across most of the galaxy is classified as Seyfert. Together with velocity dispersions exceeding 150 $\rm km\ s^{-1}$ and the symmetric velocity structure in the central region (Figure \ref{fig:accelerate}), these properties are consistent with the AGN-driven biconical ionized outflow oriented at an $\sim 26^\circ$ offset from the radio jet major axis (black dot-dashed line). Notably, the C2 component in the vicinity of the northern radio lobe reaches projected velocities up to $\rm \sim 600\ km\ s^{-1}$ and exhibits enhanced velocity dispersions of $\rm \sim 200-500\ km\ s^{-1}$ (Figure \ref{fig:accelerate}). The spatial coincidence between these extreme kinematics and the radio lobe indicates that the C2 gas in this region is dynamically influenced by the jet. The current data do not allow us to determine whether the jet is interacting with pre-existing ambient gas or with the outflowing material itself.

\begin{figure*}[ht!]
\plotone{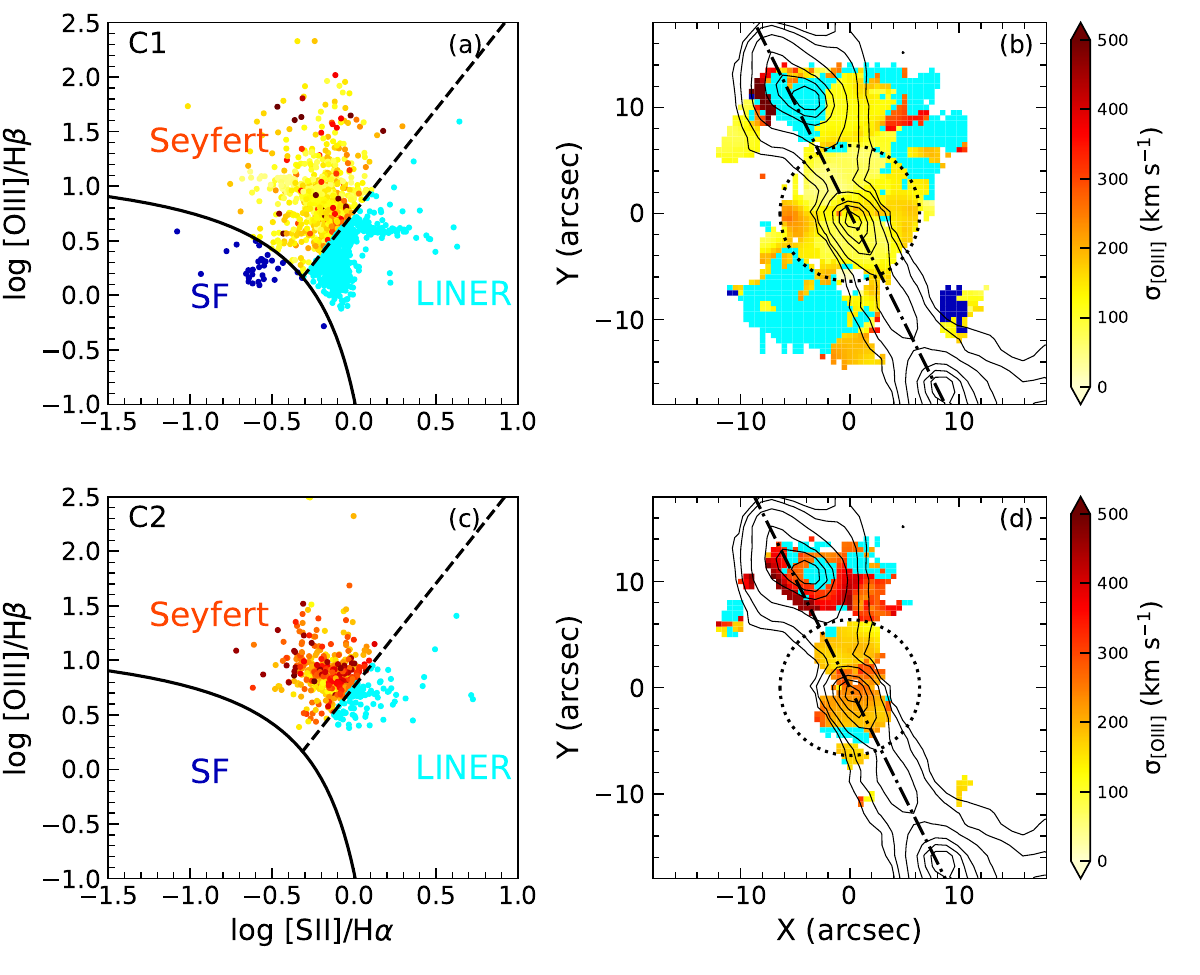} 
    \caption{Parametric \sii/\ha\ versus \oiii/\hb\ diagnostic diagrams. 
    (a) Ionization classifications for Component-1 (C1), which consists of the single Gaussian component and the narrower component from double Gaussian fitting. The solid and dashed \citep{Kewley2001,Kewley2006} curves represent the demarcation lines separating the star-forming (SF, blue), Seyfert, and LINER (cyan) regions. Seyfert spaxels are color-coded according to their \oiii\ velocity dispersion, with a color gradient from yellow to red indicating increasing velocity dispersion. 
    (b) The corresponding spatially resolved diagnostic map with color coding consistent with the corresponding diagnostic diagram shown in panel (a). Black contours indicate the VLASS 3.0 GHz radio emission and its major-axis is marked by a dot-dashed line. The dotted circle indicates the central 0.7$R_e$ region of the galaxy.
    Panels (c,d) show the same configuration as in panels (a,b), but for Component-2 (C2), corresponding to the broader component identified from double-Gaussian fitting.
    \label{fig:bpt-s2}}
\end{figure*}

Figure \ref{fig:bpt-s2} presents the \sii/\ha$-$\oiii/\hb\ diagrams for the C1 and C2 components. The classifications based on the \sii\ diagnostic are broadly consistent with those derived from the \nii\ diagram. However, the C2 component around the northern radio lobe shows a relatively stronger LINER contribution in the \sii\ diagram compared to the \nii\ diagram, further supporting the presence of shock-related ionization in this region.

\subsection{Electron Temperature} \label{sec:Te}

\begin{figure*}[ht!]
\plotone{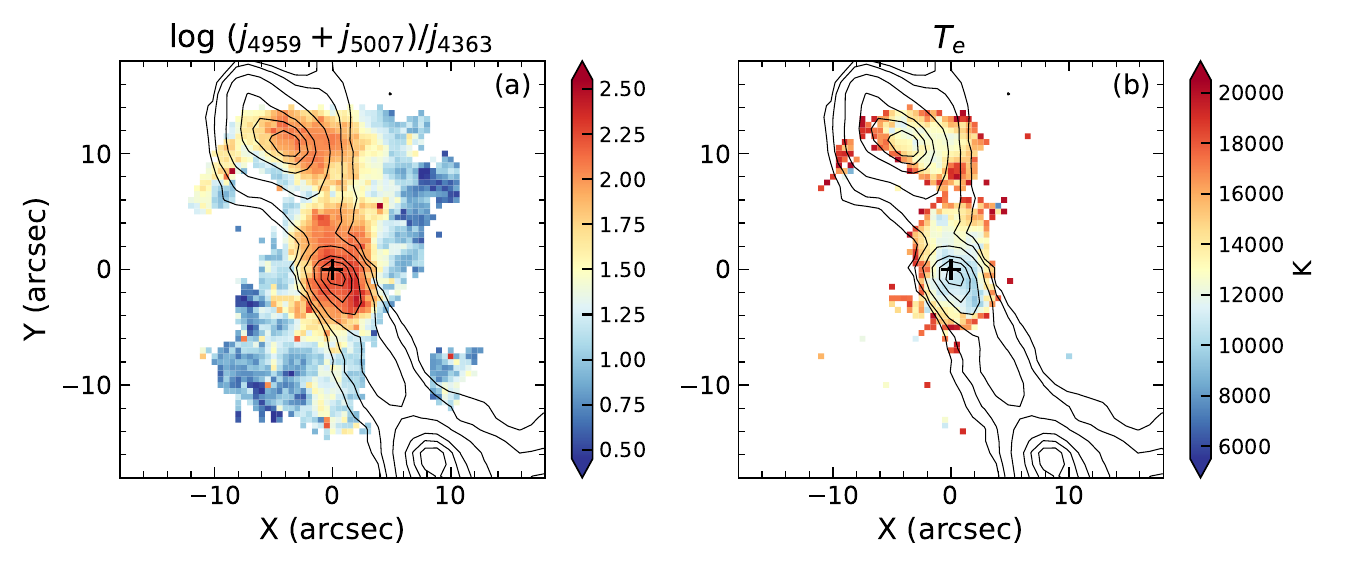} 
    \caption{Electron temperature determination.
    (a) Map of the \oiii\ emission-line ratio, defined as (\oiii$\lambda$4959+\oiii$\lambda$5007)/\oiii$\lambda$4363, and labeled in the figure as $(j_{\rm 4959}+j_{\rm 5007})/j_{\rm 4363}$. 
    (b) The spatially resolved electron temperature ($T_e$) map derived from the \oiii\ emission-line ratios following \cite{Osterbrock2006}. Black contours indicate the VLASS 3.0 GHz radio emission.
    \label{fig:Te}}
\end{figure*}

According to the ionization properties and gas kinematics around the northern radio lobe, the presence of jet-driven fast shocks is plausible and such shocks may have a significant impact on the surrounding ISM \citep{Allen2008}. Consequently, the gas temperature in this region could be enhanced by shock heating processes. To examine this possibility, the non-parametric emission-line fluxes of \oiii$\lambda$4363, \oiii$\lambda$4959, and \oiii$\lambda$5007 are used to derive the electron temperature, following the method described in \cite{Osterbrock2006}. The spatially resolved line ratio map is presented in Figure \ref{fig:Te}(a), where all spaxels have $\rm S/N>3$ for the three emission lines. Both the galaxy center and the northern radio lobe exhibit high values of the $(\oiii\lambda 5007 + \oiii\lambda 4959)/\oiii\lambda 4363$ ratio compared to the outer disk. The ratio in the radio lobe is slightly lower than that in the center, corresponding to a higher electron temperature in the lobe region. Figure \ref{fig:Te}(b) shows the electron temperature map. The electron temperature at the peak of the northern radio lobe is $\gtrsim 1000$ K higher than that in the galaxy center, which is primarily dominated by AGN photoionization. This temperature enhancement is consistent with additional shock heating likely associated with jet-ISM interaction.

\section{Discussion} \label{sec:discussion}

\subsection{Multi-channel AGN activities} \label{sec:feedback}
\subsubsection{Coexistance of Galactic Jet and Ionized Outflow} \label{sec:coexist}

Multi-wavelength observations of 4C+29.30 reveal the coexistence of a well-defined radio jet and a Seyfert-like ionized biconical outflow. A key question is how to distinguish between radiatively driven winds and jet-related gas outflows. 
Most observations of the coexistence of a radio jet and a gas outflow suggest a crucial role of the jet in the origin of the gas outflow (e.g. \mbox{\citealt{Mullaney2013}}; \citealt{Jarvis2019,Girdhar2024}). A galaxy-wide example is found in the radio-loud galaxy NGC 5972, which hosts a well-known helical-shaped extended emission-line region (EELR) extending across the stellar body. Based on radio imaging and optical ionized-gas kinematics, \cite{Ali2025} presented spatial correlations between the galactic radio jet and the ionized outflowing gas, supporting a jet-driven feedback mechanism as a key factor in the formation of the EELR. 
Such studies predominantly attribute outflows to jet-driven processes. Nevertheless, there is a lack of studies investigating the coexistence of radio jets and radiatively driven outflows in the same system. \cite{Venturi2023} analyzed the gas kinematics and energetics of the Teacup galaxy (Type 2 QSO), finding that the ionized and molecular gas outflows are likely driven by a combination of AGN radiation and the jet. 
More recently, the galactic-scale emission-line outflow from the radio-loud quasar 3C 191 has been probed \citep{Zhao2025}. There is an angular offset of $\sim40^\circ$ between the axis of \oiii\ outflow and the radio jet, ruling out jets as the dominant driver of the \oiii\ outflows. 
In 4C+29.30, the projected position angles of the galactic radio jet and ionized biconical outflow differ by $\sim$26$^\circ$. 
Using the anti-correlation between Eddington ratio ($L_{\rm bol}/L_{\rm Edd}$) and scattering fraction ($f_{\rm sc}$) shown in the literature (e.g. \citealt{Noguchi2010}), \cite{Sobolewska2012} estimated a relatively high limit of $L_{\rm bol}/L_{\rm Edd}\gtrsim 0.1$ for 4C+29.30. Therefore, the spatial misalignment between the radio jet and the biconical outflow, combined with the relatively high accretion rate of the central SMBH, disfavors a scenario in which the large-scale outflow is directly collimated by the radio jet. Instead, the outflow is more consistent with a radiatively driven wind powered by the central AGN. 

4C+29.30 is therefore consistent with a multi-channel AGN activity scenario, which can be most naturally explained by coeval but physically distinct jet and radiatively driven outflow components. Although the radio jet and the ionized outflow are launched coevally during the current AGN episode, their axes are not required to be aligned due to their different launching scales and coupling to the host galaxy ISM (e.g., \citealt{Yuan2014,Gaspari2020}). Moreover, jet orientations can evolve independently of the larger scale wind/outflow due to changes in disk alignment or black hole spin \citep{Liska2018,Liska2019,Liska2021}. 
Alternatively, the radio jet and the ionized outflow may trace different phases of AGN activity, in which a reorientation of the jet axis or a change in accretion geometry leads to the observed misalignment. More broadly, the coexistence of the galactic jet and biconical ionized outflow may be consistent with an evolutionary picture in which different feedback modes are linked to variations in the accretion state of the AGN. However, the present data do not provide sufficient physical constraints to determine whether 4C+29.30 is currently undergoing such a transition. Future numerical simulations that characterize the black hole spin and accretion state (e.g. \citealt{He2025}), together with constraints on the relative timescales of the radio jet and the AGN-driven galaxy-wide outflow will be essential for clarifying the connection between these two feedback modes.

4C+29.30 also hosts jet activity on multiple spatial scales. \cite{Liuzzo2009} presented the Very Long Baseline Interferometry (VLBI) observations of the $\rm 5\ GHz$ radio emission of 4C+29.30, revealing pc-scale jets consisting of a northern core-knot structure and a southern knot. The dynamical timescales of the northern and southern knots are estimated to be $\rm\sim15\ yr$ and $\sim70$ yr, respectively. By comparison, the spectral ages of the galactic-scale double-lobed source and the larger-scale extended radio morphology are $\lesssim33$ Myr and $\gtrsim200\rm\ Myr$, respectively \citep{Jamrozy2007}. This difference in evolutionary timescales indicates that the nuclear region hosts much younger jets, corresponding to more recent phases of black hole activity, and provides evidence for episodic AGN activities in the galaxy, as observed in other radio-loud galaxies such as Centaurus A \citep{McKinley2018} and NGC 5972 \citep{Ali2025}. The recurrent AGN activity is likely associated with the fueling of cold gas during a gas-rich merger process \citep{Chandola2010}. We will further estimate the dynamical timescale of the biconical ionized gas outflow to provide observational constraints on whether the radiative feedback arises during a transitional phase between pc-scale and galactic-scale jet activity.

\subsubsection{Jet-ISM interaction} \label{sec:jet-ISM}

In classical double radio sources, hot spots mark the working surfaces of relativistic jets, where strong shocks dissipate kinetic energy and accelerate particles to high Lorentz factors \citep{Fanaroff1974,Blandford1974}. These shocks produce intense synchrotron radiation and inject energy into the relativistic plasma that inflates large-scale radio lobes, thereby influencing the dynamics of the surrounding medium \citep{Hardcastle2007,Meisenheimer1997}.

In 4C+29.30, the northern hotspot has been detected at $\rm 0.5-2\ keV$ from Chandra observations, as exhibited in Figure \ref{fig:morph}. Spectral analysis suggest the presence of both thermal and non-thermal components \citep{Jamrozy2007}, with subsequent modeling favoring a predominantly synchrotron origin for the X-ray emission \citep{Siemiginowska2012}. The thermal component may arise from gas compressed by the expanding radio plasma, spatially coincident with regions of strong optical line emission. These properties indicate that the hotspot traces a high-pressure, particle-accelerating environment at the termination of the jet.

Such a high-pressure region is expected to drive shocks into the surrounding interstellar medium. This scenario naturally accounts for the enhanced velocity dispersions and bulk motions observed in \ha\ and \oiii\ around the northern radio lobe, supporting an interpretation in which jet-driven shocks locally dominate the ionized gas kinematics. The spatial correspondence between the hotspot and the disturbed ionized gas further strengthens the case that the jet-ISM interaction is dynamically significant in this system.

\subsubsection{Outflow and Jet Energetics} \label{sec:energy}

Assuming a biconical geometry for the galactic outflow, with one side approaching and the other receding along the line-of-sight (LOS; Figure \ref{fig:accelerate}), we can calculate the mass outflow rate and kinetic power of the ionized gas. Since that the outflow energetics within the cental $\sim 3.5'' \times 5''$ region of 4C+29.30 have been reported by \cite{Couto2020} based on the Gemini IFS observations, we apply the same method to the galactic-scale outflow for comparison. The outflow mass is estimated by
\begin{equation}
    M_{\rm out}\approx 2.3\times 10^5\ \frac{L_{41}({\rm H\alpha})}{n_3}\ M_\odot
\end{equation}
where $L_{41}({\rm H\alpha})$ is the \ha\ luminosity ($L_{\rm H\alpha}$) in units of $10^{41}\rm\ erg\ s^{-1}$, and $n_3$ is the electron density ($n_e$) in units of $10^3\rm\ cm^{-3}$. We estimate $L_{\rm H\alpha}$ from the broad emission-line component, which traces the outflowing gas, obtained through the parametric emission-line fitting (Section \ref{sec:linefit}). The dust attenuation correction is applied using the Balmer decrement, under the assumption of case B recombination \citep{Osterbrock2006} and the extinction curve of \cite{Calzetti2000}. The total $L_{\rm H\alpha}$ within the outflow region (green circle in Figure \ref{fig:accelerate}) is $\sim 5.5 \times 10^{41}\rm\ erg\ s^{-1}$. A value of $n_e \sim 100\rm\ cm^{-3}$ is adopted based on the \sii$\lambda$6718/\sii$\lambda$6732 flux ratio. The outflow mass is therefore estimated as $M_{\rm out}\sim 1.3 \times 10^7 M_\odot$, which is only slightly larger than that of the nuclear outflow derived from \cite{Couto2020} (see their Table 2), despite our data covering a substantially larger spatial extent. To verify this, we further restrict the measurement to the central $\sim 3.5'' \times 5''$ region, which corresponds to the Gemini FoV, fining that the outflow mass is comparable to that reported by \cite{Couto2020}. This consistency supports the reliability of our estimation method. The small difference between the galaxy-scale and nuclear outflow masses implies that the bulk of the outflowing gas could be confined to the nuclear region of the galaxy.

The mass outflow rate can be measured as
\begin{equation}
    \dot{M}=1.4\ n_e\ m_p\ v_{\rm out}\ A\ f
\end{equation}
where $m_p=1.7 \times 10^{-24}\rm\ g$ is the proton mass. The factor 1.4 is applied to account for elements heavier than hydrogen. $v_{\rm out}$ represents the intrinsic outflow velocity derived from the line-of-sight velocity ($v_{\rm los}$) and the cone inclination angle ($\beta$), with $v_{\rm out}=v_{\rm los}/{\rm sin}\ \beta$. $\beta$ is defined as the angle between the bicone axis and the plane of the sky. We adopt $\beta=40^\circ$, following \cite{Couto2020}, who argued that the cone inclination is unlikely to differ significantly from that of the radio jet \citep{Liuzzo2009}. $A$ is the cross-sectional area of the cone. The filling factor can be calculated assuming case B recombination \citep{Osterbrock2006}
\begin{equation}
    f = 2.6 \times 10^{59}\ \frac{L_{41}({\rm H\alpha})}{V\ n_3^2}
\end{equation}
with $V=Ah/3$ denoting the volume of the outflow cone in units of $\rm cm^3$, where $h$ is the cone height derived from the projected size ($R_{\rm proj}$) of the outflow region as $h=R_{\rm proj}/{\rm cos}\ \beta$. The mass outflow rate of the galactic ionized outflow can be estimated as $\dot{M}_{\rm out}\sim 1.8\ M_\odot\ yr^{-1}$ with an average $v_{\rm out}\sim 175\rm\ km\ s^{-1}$. 
We subsequently combine the bolometric luminosity of the nucleus of 4C+29.30 ($L_{\rm bol}\sim 10^{45}\rm\ erg\ s^{-1}$; \citealt{Siemiginowska2012}) and the maximum outflow velocity ($v_{\rm out} \sim\ 717\rm\ km\ s^{-1}$) to further assess whether the radiation-driven winds can move that much material. The mass rate of a radiation-driven wind in the outflow region is $\dot{M}_{\rm rad} = L_{\rm bol}/cv_{\rm out}\sim 7.4\ M_\odot\ yr^{-1}$, which is sufficient to drive the galactic ionized gas outflows.

We then measure the kinetic power of the galactic outflow based on the established method \citep{Holt2006,Mahony2016}:
\begin{equation}
    \dot{E} = 6.34 \times 10^{35}\ \frac{\dot{M}_{\rm out}}{2} (v_{\rm out}^2 + 3\sigma_{\rm out}^2)
\end{equation}
where $\sigma_{\rm out}$ is the velocity dispersion of the broad emission-line component within the outflow region. The kinetic power is $\dot{E}_{\rm out}\sim 9.5\ \times 10^{40}\rm\ erg\ s^{-1}$, derived using the average values of $v_{\rm out}\sim 175\rm\ km\ s^{-1}$ and $\sigma_{\rm out}\sim 213\rm\ km\ s^{-1}$.

For the accelerated region around the northern radio lobe (red sector in Figure \ref{fig:accelerate}), we estimate the energetics by simply assuming a partially conical geometry with an inclination of $40^\circ$, consistent with the radio jet. Using the equations above, we calculate the accelerated gas mass to be $M_{\rm acc}\sim 4.6 \times 10^6 M_\odot$ from the total $L_{\rm H\alpha}$ within the accelerated region, which is $\sim 2.0 \times 10^{41}\rm\ erg\ s^{-1}$. Based on the average $v_{\rm out}\sim\rm\ 573\ km\ s^{-1}$ and $\sigma_{\rm out}\sim\rm\ 342\ km\ s^{-1}$ within this region, the mass outflow rate and the kinetic power are estimated as $\dot{M}_{\rm acc}\sim 0.87\ M_\odot\ yr^{-1}$ and $\dot{E}_{\rm acc}\sim 1.9 \times 10^{41}\rm\ erg\ s^{-1}$, respectively. Considering that the lower limit to the jet kinetic energy required to power the radio lobes is $\sim 10^{42}\rm\ erg\ s^{-1}$ \citep{Siemiginowska2012}, the jet is energetically capable of driving the observed gas motions around the northern radio lobe.

\subsection{Does the Radio Jet Influence Star Formation?} \label{sec:starform}

\subsubsection{Mass Assembly History} \label{sec:assembly}
Given the clear kinematic and ionization signatures of jet-ISM interaction in the northern radio lobe, it is important to assess whether the stellar population in this region shows any evidence potentially linked to the jet activity. We therefore examine the spatially resolved stellar mass assembly history derived from the MaNGA Pipe3D data.

\begin{figure*}[ht!]
    \centering
    \includegraphics[width=\textwidth]{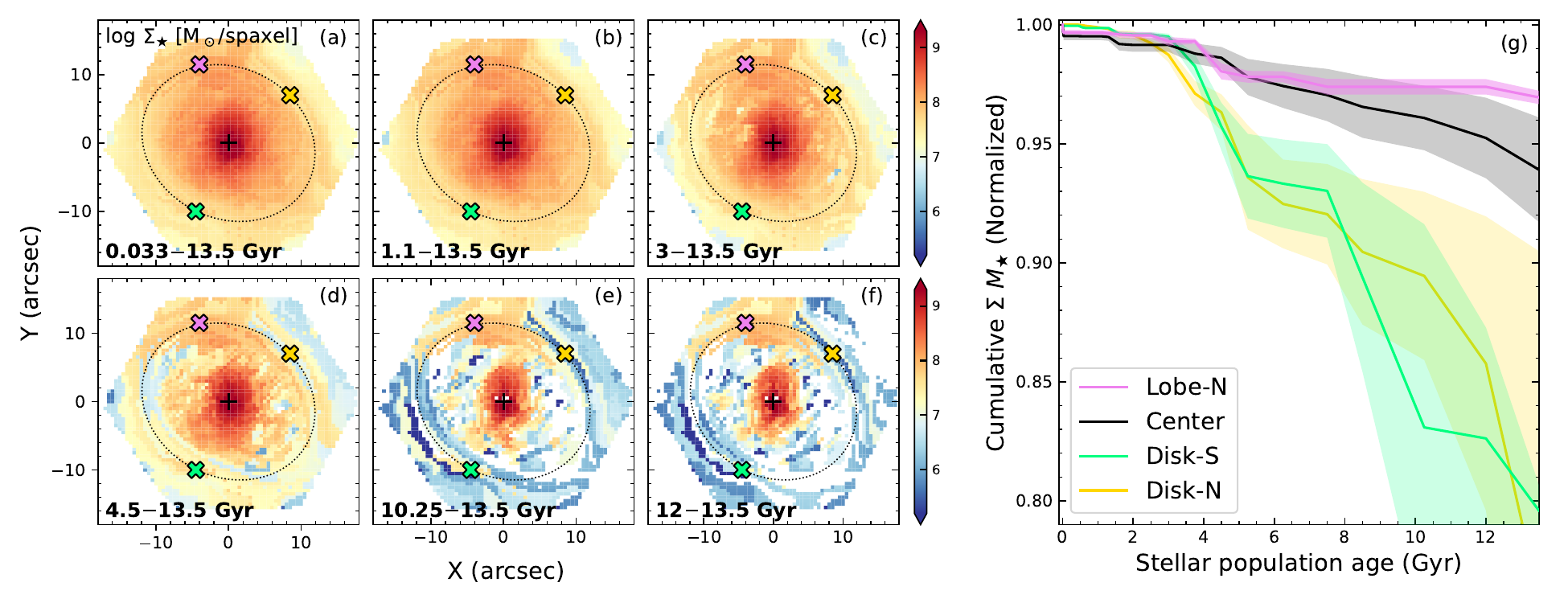}
    \caption{Mass assembly history. 
    (a-f) Stellar-mass surface density ($\Sigma_\star$) maps derived from the MaNGA Pipe3D data for different stellar age ranges, as indicated in the bottom-left insets. The dotted ellipse, centered on the galaxy photometric center (black cross, Center), encompasses the flux peak (magenta cross, Lobe-N) of the northern radio lobe. Its geometric parameters are defined by the optical morphology of the galaxy. The yellow (Disk-N) and green (Disk-S) crosses mark the northern and southern disk regions, respectively; each is located at the same galactocentric radius as the radio flux peak. 
    (g) The cumulative $\Sigma_\star$ as a function of stellar age for the positions shown in panels (a-f). Solid lines represent the cumulative $\Sigma_\star$ curves, corresponding to the median $\Sigma_\star$ of spaxels within a 3$''$ radius aperture centered on each marked position. Shaded regions indicate the 1$\sigma$ scatter of $\Sigma_\star$. 
    \label{fig:growth}}
\end{figure*}

According to the methods established in \cite{Sanchez2022}, the stellar mass in each spaxel is computed by combining the mass-to-light ratio of each simple stellar population (SSP) with its corresponding light fraction within different stellar age bins. Figure \ref{fig:growth}(a-f) presents stellar mass surface density ($\Sigma_\star$) maps in six representative stellar age intervals. The magenta cross marks the flux peak of the northern radio lobe (Lobe-N), while comparison regions at similar galactocentric radii in the northern and southern disks are indicated by yellow (Disk-N) and green (Disk-S) crosses. At fixed radius, the $\Sigma_\star$ in the Lobe-N region is systematically higher than that in the disk regions for the oldest stellar populations. 

The cumulative $\Sigma_\star$ as a function of stellar age (Figure \ref{fig:growth}g), measured as the median value of spaxels within a 3$''$ aperture around each region, further shows that the Lobe-N region experienced an early and rapid mass assembly compared to the disk and central regions. Approximately 97\% of the stellar mass in Lobe-N was formed several Gyr ago, significantly longer than the typical radio jet lifetimes of $10^6-10^7$ yr. This timescale mismatch indicates that the dominant quenching event in this region must have occurred long before the current episode of jet activity. Instead, the mass assembly history suggests an early buildup followed by long-term quiescence. The irregular optical morphology observed around the northern radio lobe further supports a scenario in which the old stellar component is associated with merger-remnant structures rather than recent jet-driven processes.

\subsubsection{Metallicity and Age Indicators} \label{sec:stellar}
\begin{figure*}[ht!]
\plotone{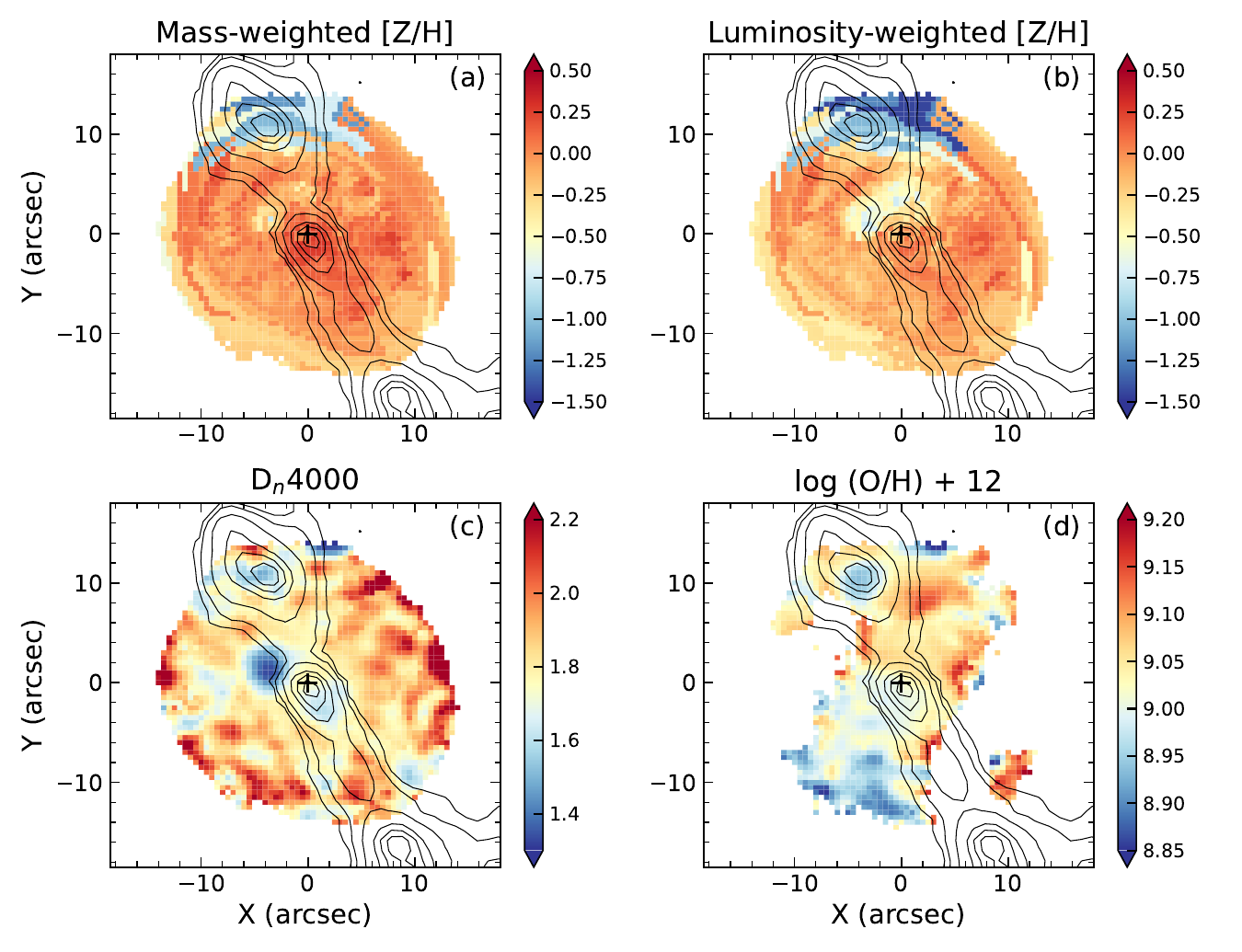} 
    \caption{Stellar populations and gas-phase properties. 
    (a,b) Mass- and luminosity-weighted stellar metallicities ([Z/H]) derived from the MaNGA Pipe3D data.
    (c) Spatially resolved D$_n$4000 map.
    (d) Spatially resolved gas-phase metallicity derived using the \nii/\sii\ line ratio as the indicator. Black contours indicate the VLASS 3.0 GHz radio emission.
    \label{fig:sf}}
\end{figure*}

Figure \ref{fig:sf}(a,b) present the mass- and luminosity-weighted stellar metallicity ([Z/H]) maps derived from the MaNGA Pipe3D data, with radio contours overlaid for reference. The region surrounding the northern radio lobe exhibits a systematically lower mass-weighted metallicity compared to the rest of the galaxy. This suggests that the stellar mass in this region is dominated by an evolved population with comparatively lower metallicity, consistent with the early and rapid mass assembly inferred in Section \ref{sec:assembly}.
The luminosity-weighted metallicity is similarly low and is lower than the mass-weighted value by $\sim$0.1 dex. This indicates that the stellar light is slightly biased toward a younger and/or relatively metal-poor population. Nevertheless, the small offset implies that this population represents only a minor fraction of the total stellar mass and luminosity, inconsistent with a dominant recent starburst.

We further examine the resolved D$_n$4000 map (Figure \ref{fig:sf}c), which traces the luminosity-weighted stellar age. A decrease in D$_n$4000 is observed around the northern radio lobe relative to adjacent regions, indicating the presence of a younger stellar component. However, when considered together with the stellar metallicity diagnostics, this decrease likely reflects only a modest contribution from younger and/or lower-metallicity stars superimposed on an evolved stellar population. The gas-phase metallicity (Figure \ref{fig:sf}d), traced by the \nii/\sii\ emission-line ratio, is also reduced in the northern lobe region, indicating a chemically less enriched interstellar medium. If ongoing star formation is present, it could produce relatively metal-poor young stars consistent with the observed luminosity-weighted metallicity. The low-metallicity gas could be related to merger-driven redistribution or external accretion.

In addition to the northern radio lobe, a region near the galaxy center exhibits even lower D$_n$4000 values. This region coincides with a distinct optical counterpart, but does not show corresponding features in either the gas kinematics or ionization diagnostics. Although it lies close to the projected edge of the radio jet, no clear signatures of jet-ISM interaction are detected. Together, these properties suggest that the low D$_n$4000 is likely associated with a localized young stellar population in the disk, rather than being directly linked to the radio jet activity.

\subsubsection{Possible Origin of NUV Emission} \label{sec:nuv}

\begin{figure*}[ht!]
    \centering
    \includegraphics[width=\textwidth]{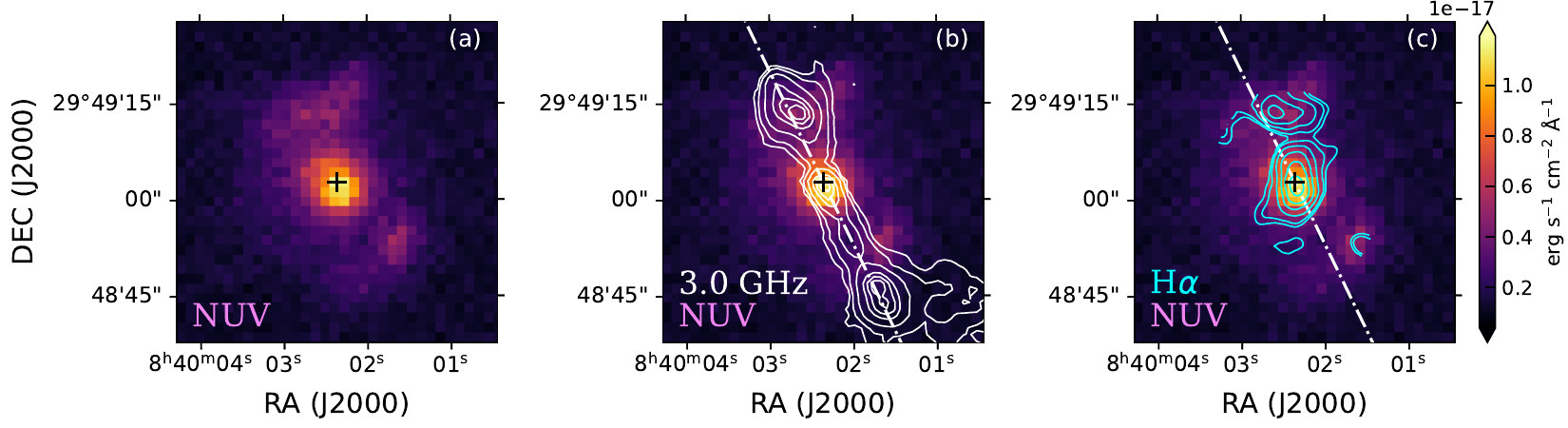}
    \caption{Spatial comparison of the NUV emission, radio jet and \ha\ morphology. 
    (a) GALEX NUV image.
    (b) White contours show the VLASS 3.0 GHz radio emission overlaid on the NUV image. 
    (c) Cyan contours show the non-parametric \ha\ flux from DAP overlaid on the NUV image.  
    The white dot-dashed line in panels (b,c) marks the major axis of the radio jet.
    \label{fig:nuv}}
\end{figure*}

Since D$_n$4000 traces stellar populations on intermediate timescales, we further examine the ultraviolet (UV) imaging to probe more recent star formation and its potential connection to the radio jet.
UV emission primarily traces massive O-, B-, and A-type stars and is therefore sensitive to star formation on timescales of $\rm 10^6-10^8$ yr. Spatial alignment between radio jets and UV emission has been reported in several systems and is often interpreted as evidence for jet-triggered star formation \citep{Drouart2016,Zirm2005,Rubinur2024}. A well-known nearby example is Centaurus A, where UV-bright star-forming regions are preferentially distributed along the jet direction \citep{Joseph2022a}. The near-ultraviolet (NUV) emission of 4C+29.30 is detected in archival observations from Galaxy Evolution Explorer (GALEX; \citealt{Martin2005}). We stack two exposures (totaling 21 ks) to improve the signal-to-noise ratio. As shown in Figure \ref{fig:nuv}, enhanced NUV emission is observed around the northern radio lobe, with weaker enhancement near the boundary of the southern lobe. The NUV morphology is also spatially coincident with the non-parametric \ha\ flux distribution.

The spatial correspondence between NUV emission and the radio structure is suggestive of localized star formation that may be associated with the jet. However, given the disturbed optical morphology and merger-remnant features of the system, merger-induced star formation cannot be excluded. Assuming that the NUV emission is entirely powered by young stars, we estimate an average SFR surface density of $\rm 0.02\pm 0.004\ M_\odot\ yr^{-1}\ kpc^{-2}$ and $\rm 0.01\pm 0.003\ M_\odot\ yr^{-1}\ kpc^{-2}$ for the northern and southern radio lobes, respectively, following the method and calibration of \cite{Joseph2022b}. These values are $\sim$1 dex elevated compared to regions at similar galactocentric radii without radio emission. Nevertheless, UV emission can also arise from AGN-related processes, including scattered nuclear continuum and nebular continuum emission \citep{Pentericci1999,Allen2002}. Additional constraints, such as UV polarization measurements (e.g. \citealt{Pentericci1999}) and sensitive cold-gas observations (e.g. \citealt{Oosterloo2005,Croft2006,Salome2015}), are required to robustly determine whether ongoing star formation is directly triggered by the jet.

Overall, 4C+29.30 presents a system in which the central AGN simultaneously drives both a radiative wind and a mechanical jet, illustrating the coexistence of multiple feedback channels. The northern radio lobe locally dominates the ionized gas kinematics and ionization structure, consistent with fast shocks associated with jet-ISM interaction. The stellar population in this region is primarily old and predates the current AGN episode. Any recent star formation, if present, appears modest and cannot be unambiguously attributed to the jet.

\section{Conclusions} \label{sec:conclusion}

In this paper, we combined optical IFS data from SDSS/MaNGA and CFHT/SITELLE to investigate the gas kinematics and ionization properties of the post-merger galaxy 4C+29.30. This system harbors an active SMBH and exhibits prominent radio structures, including a core, jets, and double lobes, as revealed in the VLASS 3.0 GHz image. We examine the physical connections between the central AGN and the host galaxy by analyzing signatures of AGN activity and correlations between the radio jet and the ISM. Our main findings are summarized as follows:

\begin{enumerate}
    \item The stellar and gaseous velocity fields reveal a rotating gas component whose kinematic position angle is offset by $87^\circ$ from that of the stellar component. Such a strong misalignment is likely associated with a gas-rich merger, during which the angular momentum direction of the pre-existing gas was redistributed, leading to the observed decoupling between gas and stars.
    \item A Seyfert-like biconical ionized outflow is identified from the gaseous kinematics, with its axis misaligned by $\sim 26^\circ$ relative to the major axis of the radio jet. A radiation-driven wind with a mass rate of $\sim 7.4\ M_\odot\ yr^{-1}$ is sufficient to drive such galactic-scale biconical outflowing gas whose mass outflow rate is $\sim 1.8\ M_\odot\ yr^{-1}$. The coexistence of the radio jet and a radiatively driven outflow in 4C+29.30 supports a multi-channel feedback scenario, in which distinct AGN feedback mechanisms operate concurrently. Alternatively, the observed structures may reflect different phases of AGN activity, dominated at different times by jet-driven or radiatively driven processes. Future numerical simulations constraining the black hole spin and accretion state will be essential for distinguishing between these scenarios.
    \item A region around the northern radio lobe exhibits additional non-circular gas motions, characterized by enhanced velocities and velocity dispersions relative to the rest of the galaxy. This region shows contributions from both Seyfert and LINER ionization. The jet kinetic power that feeds the radio lobes ($\rm\sim 10^{42}\ erg\ s^{-1}$) is larger than the kinetic power of the accelerated gas in this region ($\rm\sim 1.9\times 10^{41}\ erg\ s^{-1}$). We suggest that fast shocks, likely driven by interactions between the expanding radio lobe and the ISM, play a key role in shaping both the gas kinematics and ionization structure.
\end{enumerate}

4C+29.30 therefore provides a representative laboratory for studying multi-channel AGN feedback in a post-merger environment, with the stellar population properties in the northern radio lobe offering complementary clues to its local impact. The stellar mass assembly history and chemical abundances indicate that this region is dominated by an old population likely associated with merger remnants rather than recent AGN-driven quenching. At the same time, the locally reduced gas-phase metallicity, lower D$_n$4000 index, and spatial alignment of the NUV and \ha\ emission with the radio structure suggest a modest contribution from younger stars superimposed on the evolved component. Follow-up CO observations with higher spatial and spectral resolution will be essential for clarifying the role of the jet in regulating star formation.

\begin{acknowledgments}

    We thank the anonymous referee for helpful comments that improved this paper.
    We thank Dr. Feng Yuan, Dr. Suoqing Ji, Dr. Yijun Wang, Dr. Zhenya Zheng, Dr. Benoît Epinat, and Dr. Si-Yue Yu for helpful discussions.
    We acknowledges the National Key R\&D Program of China (Grant NO.2023YFA1607904), the China Postdoctoral Science Foundation under Grant Number 2025M773188, the Postdoctoral Fellowship Program of CPSF under Grant Number GZB20250734, the National Natural Science Foundation of China (NSFC) grants 12333002, 12221003, the China Manned Space Project with No. CMS-CSST-2025-A10 and CMS-CSST-2025-A07, and the National SKA Program of China No. 2025SKA0150103. 
    Y.M.C acknowledges support by the National Natural Science Foundation of China, NSFC Grant No. 12333002 and the China Manned Space Project with No. CMS-CSST-2025-A08.
    M.B. acknowledges support by the National Natural Science Foundation of China, NSFC Grant No. 12303009.
    X.X. acknowledges the NSFC grant 12403018, the China Postdoctoral Science Foundation (No. 2023M741639), and the Jiangsu Funding Program for Excellent Postdoctoral Talent (No. 2024ZB249).
    T.F. is supported by the National SKA Program of China No. 2025SKA0150103, National Natural Science Foundation of China under Nos. 12550002, 12133008, 12221003, 11890692, and the science research grants from the China Manned Space Project with No. CMS-CSST-2021-A04 and No. CMS-CSST-2025-A10.
    J.W. acknowledges the National Key R\&D Program of China (Grant NO.2023YFA1607904) and the National Natural Science Foundation of China (NSFC) grants 12221003.

    Funding for the Sloan Digital Sky 
    Survey IV has been provided by the 
    Alfred P. Sloan Foundation, the U.S. 
    Department of Energy Office of 
    Science, and the Participating 
    Institutions. 

    SDSS-IV acknowledges support and 
    resources from the Center for High 
    Performance Computing  at the 
    University of Utah. 
    The SDSS 
    website is www.sdss.org.
    SDSS-IV is managed by the 
    Astrophysical Research Consortium 
    for the Participating Institutions 
    of the SDSS Collaboration including 
    the Brazilian Participation Group, 
    the Carnegie Institution for Science, Carnegie Mellon University, Center for 
    Astrophysics | Harvard \& 
    Smithsonian, the Chilean Participation 
    Group, the French Participation Group, 
    Instituto de Astrof\'isica de 
    Canarias, The Johns Hopkins 
    University, Kavli Institute for the 
    Physics and Mathematics of the 
    Universe (IPMU) / University of 
    Tokyo, the Korean Participation Group, 
    Lawrence Berkeley National Laboratory, 
    Leibniz Institut f\"ur Astrophysik 
    Potsdam (AIP),  Max-Planck-Institut 
    f\"ur Astronomie (MPIA Heidelberg), 
    Max-Planck-Institut f\"ur 
    Astrophysik (MPA Garching), 
    Max-Planck-Institut f\"ur 
    Extraterrestrische Physik (MPE), 
    National Astronomical Observatories of 
    China, New Mexico State University, 
    New York University, University of 
    Notre Dame, Observat\'ario 
    Nacional / MCTI, The Ohio State 
    University, Pennsylvania State 
    University, Shanghai 
    Astronomical Observatory, United 
    Kingdom Participation Group, 
    Universidad Nacional Aut\'onoma 
    de M\'exico, University of Arizona, 
    University of Colorado Boulder, 
    University of Oxford, University of 
    Portsmouth, University of Utah, 
    University of Virginia, University 
    of Washington, University of 
    Wisconsin, Vanderbilt University, 
    and Yale University.

    Our SITELLE observation (25AS007) is kindly supported by China
    Telescope Access Program (TAP). 
    Based on observations obtained at the Canada-France-Hawai`i Telescope (CFHT) which is operated by the National Research Council of Canada, the Institut National des Sciences de l'Univers of the Centre National de la Recherche Scientifique of France, and the University of Hawai`i. CFHT is located on Maunakea on Hawai`i Island, a mountain of considerable cultural, natural, and ecological significance. Maunakea is a sacred site to Native Hawaiians, also known as Kānaka `Ōiwi. Quality observations are made possible by relentless effort of the entire staff at Canada-France-Hawai`i Telescope. Based on observations obtained with SITELLE, a joint project between Universite Laval, ABB-Bomem, Universite de Montreal, and the CFHT with funding support from the Canada Foundation for Innovation (CFI), the National Sciences and Engineering Research Council of Canada (NSERC), Fond de Recheche du Quebec - Nature et Technologies (FRQNT) and CFHT.

\end{acknowledgments}

The optical HST and NUV GALEX data presented in this article were obtained from the Mikulski Archive for Space Telescopes (MAST) at the Space Telescope Science Institute. The specific observations analyzed can be accessed via \dataset[DOI: 10.17909/g2n6-hh92]{https://doi.org/10.17909/g2n6-hh92}

\bibliography{4C2930-paper-xcao}{}
\bibliographystyle{aasjournalv7}

\end{document}